\documentclass[12pt,dvips]{article}

\usepackage{graphicx}
\usepackage{rotating}
\usepackage{axodraw}
\usepackage{epsfig}
\usepackage{color}

\newcommand{\imag}{\Im {\rm m}}
\newcommand{\real}{\Re {\rm e}}

\newcommand{\stau}{{\tilde{\tau}}}
\newcommand{\lsim}{\mathrel{\mathop{\kern 0pt \rlap
  {\raise.2ex\hbox{$<$}}}
  \lower.9ex\hbox{\kern-.190em $\sim$}}}
\newcommand{\gsim}{\mathrel{\mathop{\kern 0pt \rlap
  {\raise.2ex\hbox{$>$}}}
  \lower.9ex\hbox{\kern-.190em $\sim$}}}
\newcommand{\beq}     {\begin{equation}}
\newcommand{\eeq}     {\end{equation}}
\newcommand{\bea}     {\begin{eqnarray}}
\newcommand{\eea}     {\end{eqnarray}}

\newcommand{\A}       {{\mathcal A}}

\newcommand{\no}      {\nonumber}
\newcommand{\thin}      {\hspace{.5mm}}
\newcommand{\neu}      {{\tilde{\chi}^0}}

\begin{document}

\renewcommand{\thefootnote}{\fnsymbol{footnote}}

\mbox{ }\\[-1cm]
\mbox{ }\hfill KIAS--P03078\\[-1mm]
\mbox{ }\hfill TUM--HEP--529/03\\[-1mm]
\mbox{ }\hfill \today

\bigskip
\bigskip
\bigskip

\begin{center}
{\Large \bf Analysis of CP Violation in Neutralino Decays to Tau
Sleptons} \\[1cm]
{\sc S.Y. Choi$^{1}$,\, M, Drees$^{2}$,\, B. Gaissmaier$^{2}$ and
J. Song$^{3}$} 
\end{center}

\bigskip

{\small
\begin{enumerate}
\item[{}] $^1${\it Department of Physics, Chonbuk National
                   University, Chonju 561--756, Korea}\\[-0.7cm]
\item[{}] $^2${\it Physik Department, TU M\"{u}nchen, James
                   Franck Str., 85748 Garching, Germany}\\[-0.7cm]
\item[{}] $^3${\it Department of Physics, Konkuk University,
                   Seoul 143-701, Korea}
\end{enumerate}
}

\bigskip
\bigskip
\bigskip

\begin{abstract}
\noindent In the minimal supersymmetric standard model, tau sleptons
$\tilde \tau_{1,2}$ and neutralinos $\tilde \chi^0_{1,2}$ are expected
to be among the lightest supersymmetric particles that can be produced
copiously at future $e^+e^-$ linear colliders. We analyze $\tilde
\tau$ pair and $\tilde \chi_1^0 \tilde \chi_2^0$ production under the
assumption $m_{\tilde \chi_1^0} < m_{\tilde \tau_1} < m_{\tilde
\chi_2^0}$, allowing the relevant parameters of the SUSY Lagrangian to
have complex phases. We show that the transverse and normal components
of the polarization vector of the $\tau$ lepton produced in $\tilde
\chi_2^0$ decays offer sensitive probes of these phases.
\end{abstract}

\newpage

\section{Introduction}
\label{sec:Introduction}

In extending the standard model (SM) of strong and electroweak
interactions, supersymmetry (SUSY) provides a well--motivated
framework with several virtues~\cite{Nilles_Haber}.  Weak--scale SUSY
has its natural answer to the gauge hierarchy
problem~\cite{Witten:nf}, achieves gauge unification without the ad
hoc introduction of additional particles~\cite{Unification}, and
radiatively explains the electroweak symmetry breaking in terms of the
large top quark mass~\cite{Ibanez:fr}. Moreover, $R$ parity conserving
SUSY offers a compelling candidate for the cold dark-matter component
of the Universe~\cite{Ellis:1983ew}.

Since none of the superpartners has been found, SUSY is apparently
broken if it exists at all. Only {\em soft} breaking terms are allowed,
however, if SUSY is meant to solve the hierarchy problem. Even though
this softness leads to experimentally accessible signatures of the
superpartners, the presence of many new (generally complex) parameters
complicates the phenomenological situation: Without a specific
mechanism for SUSY breaking, even in the minimal supersymmetric
standard model (MSSM) more than one hundred new parameters are
introduced~\cite{Haber}. Therefore precision measurements of these
soft breaking terms as well as other fundamental SUSY parameters are
essential to explore the SUSY breaking mechanism. Here we discuss how
to extract the parameters, in particular the CP--odd phases, in the
scalar tau (stau) and neutralino sectors of the MSSM. Recall that
CP--odd phases are needed~\cite{Sakharov} for the dynamical generation
of the baryon asymmetry of the Universe; the phase in the CKM (quark
mixing) matrix is most likely not sufficient.

The third generation scalars' {\em CP} phases have drawn a lot of
interest as these phases can be large without any conflict with
experimental constraints. In contrast, the first two generation
sfermions are severely constrained by measurements of the electric
dipole moments of the electron, muon, and neutron \cite{pdg}: Either
their {\em CP} violating phases are very small or their masses are
very large, well above 1 TeV, or parameters are adjusted such that
different contributions cancel to very good precision~\cite{cancel}.
Apart from the absence of a well-established mechanism to suppress the
{\em CP} violating phases for the first two generations, other
difficulties such as potentially large flavor changing neutral
currents~\cite{Gabbiani:1996hi}, and rapid proton decay through
dimension five operators in SUSY GUTs~\cite{Goto:1998qg}, prefer very
heavy first two generation sfermions~\cite{Bagger:1999sy}. In
contrast, third generation sfermions are expected to be light
(``inverted hierarchy'') due to their central role in the naturalness
problem as well as due to their large Yukawa couplings which
substantially reduce their masses at the electroweak scale. Therefore,
${\cal O}(1)$ {\em CP} violating phases and relatively small masses of
third generation sfermions are well motivated and phenomenologically
viable.

The stau sector may well allow the first measurement of {\em CP}
violating phases of sfermions at a future linear collider (FLC) with
c.m.~energy of up to 500 GeV: In inverted hierarchy models, only some
light neutralinos, charginos and third--generation sfermions are
expected to be produced at the FLC whereas the other supersymmetric
particles are not accessible~\cite{SNOWMASS}. Our main focus is on the
{\em CP} violating phase in the stau sector, $\phi_\stau$. This phase
is associated with $\tilde \tau_L - \tilde \tau_R$ mixing, which in
turn is enhanced for large values of the ratio $\tan\beta$ of Higgs
vacuum expectation values. Note that $\tan\beta$ as high as $ 50$ is
preferred in some $SO(10)$ grand unified models with Yukawa
unification~\cite{LargeTanB}. In contrast, scenarios with $\tan\beta$
near unity are severely constrained by Higgs searches at LEP
\cite{lephiggs}. We will see that $\tan\beta = 10$ is sufficient to
generate large {\em CP}--violating effects.

A phenomenologically significant question is which observable is
sensitive to the phase $\phi_\stau$.  The stau is expected to be
advantageous in constructing {\em CP}-odd observables since one of its
decay products, the tau lepton, also decays, which enables us to
measure the tau polarization.  In the simplest decay channel
$\stau^\pm \to \tau^\pm \neu_1$, however, the invisibility of the LSP
$\neu_1$ allows only two measurable vectors, the three-momentum and
polarization vector of the tau: No {\em CP}--odd observable can be
constructed.  Among two-step cascade decays, the decay of the second
lightest neutralino $\neu_2$ into a stau and a tau lepton, followed by
the stau decay, is one of the most promising candidates especially if
$\tan\beta$ is not small. $\tilde \chi_1^0 \tilde \chi_2^0$ production
has one of the lowest threshold energies of all SUSY processes that
lead to visible final states. Moreover, if $m_{\tilde \chi_1^0} <
m_{\tilde \tau_1} < m_{\tilde \chi_2^0}$, the decay $\tilde \chi_2^0
\rightarrow \tilde \tau_1^\pm \tau^\mp$ has a large branching ratio,
often near 100\%. Finally, {\em CP}--odd observables\footnote{In this
paper we do not distinguish between {\em CP} and {\em T} violation,
since we assume the {\em CPT} theorem to hold.} can be constructed,
since the $\tilde \chi_2^0$ in the intermediate state is
polarized. Previous studies \cite{oldcp} have focused on $\chi_2^0
\rightarrow \ell^+ \ell^- \tilde \chi_1^0$ decays $(\ell = e, \, \mu,
\, \tau)$, using the triple product of the $\ell^\pm$ 3--momenta with
the incident $e^-$ beam 3--momentum as {\em CP}--odd observable. This
allows to probe {\em CP} violation in the neutralino sector, but is
{\em not} sensitive to $\phi_{\tilde \tau}$. Here we focus on the case
$\ell = \tau$, and construct {\em CP}--odd observables involving the
spin of of the $\tau$ lepton produced in the first step of $\tilde
\chi_2^0$ decay.\footnote{The process $e^+e^- \rightarrow \tilde
\chi_1^0 \tilde \chi_2^0 \rightarrow \tilde \chi_1^0 \tilde \chi_1^0
\tau^+ \tau^-$ has very recently also been studied in
ref.\cite{bartltau}. In that study identical soft breaking masses in
the $\tilde e$ and $\tilde \tau$ sectors were assumed. This leads to
much larger signal cross sections; however, the experimental bounds on
$d_e$, which were not considered in ref.\cite{bartltau}, greatly
constrain such scenarios.}

Our purpose in this paper is to show that these observables are indeed
sensitive to $\phi_{\tilde \tau}$. We work in the general {\em
CP}--noninvariant MSSM framework with sizable $\tan\beta$ and heavy
first and second generation sfermions. A specific {\em CP}--violating
scenario for SUSY parameters is introduced in
Sec.~\ref{sec:Scenario}. In Sec.~\ref{sec:Mixing} we describe the
mixing formalism in the stau and neutralino sectors, focusing on their
{\em CP} properties. Section \ref{sec:Production} deals with the cross
sections for the neutralino pair and the stau pair production at an
$e^+ e^-$ collider with polarized beams. The assumption of heavy first
and second generation sfermions simplifies the helicity amplitudes of
the process $e^+ e^- \to \neu_i \neu_j$.  The expression for the
neutralino polarization vector is also given.  Section \ref{sec:Decay}
is devoted to the formal description of the decays of the stau and the
neutralino with special emphasis on the polarization of the tau
leptons. Through the spin/momentum correlations of the neutralino and
tau lepton, we suggest that the polarization asymmetries of the final
tau lepton are useful to probe the $CP$ violating phase. Sec.~6
contains a Monte Carlo simulation of two points of parameter space. We
find {\em CP}--odd asymmetries of up to 30\% for some regions of phase
space. Finally, Sec.~\ref{sec:Conclusions} summarizes our results and
concludes.

\section{A SUSY scenario}
\label{sec:Scenario}

As a general framework, we consider the {\em CP}--noninvariant MSSM
with sizable $\tan\beta$.  We also assume that the first and second
generation sfermions are heavy enough to be effectively decoupled from
the theory for a 500 GeV LC. A large {\em CP} violating phase in the
stau sector is then allowed without violating any constraint from the
electric dipole moments of the electron, neutron and mercury atom.  We
choose the c.m. energy of the FLC such that $\tilde \chi_1^0 \tilde
\chi_2^0$ production is possible whereas $\tilde \chi_1^0 \tilde
\chi_3^0$ production is not. We also assume that $e^+e^- \rightarrow
\tilde \tau_1^\pm \tilde \tau^\mp_{1,2}$ production can be studied,
possibly by running the FLC at a higher energy. We assume that the
lightest supersymmetric particle (LSP) is the lightest neutralino
$\tilde{\chi}^0_1$. Because of $R$--parity conservation, the LSP is
stable and escapes detection. The decay products of any SUSY particle
contains at least one $\tilde{\chi}^0_1$.

One of the core missions of the FLC is the precision measurement of
the fundamental SUSY parameters, the first step to reveal the
supersymmetry breaking mechanism and to open a window onto the final
theory at the Planck scale~\cite{Blair:2002pg}. The
chargino/neutralino sector involves as fundamental parameters the
$SU(2)$ gaugino mass $M_2$ (which can be chosen real and positive),
the $U(1)$ gaugino mass $M_1=|M_1|\, {\rm e}^{i\Phi_1}$, the higgsino
mass parameter $\mu=|\mu|\, {\rm e}^{i\Phi_\mu}$, and the parameter
$\tan\beta$. The stau sector has the $SU(2)$ doublet and singlet soft
breaking mass parameters $\widetilde{m}_L$ and $\widetilde{m}_R$, the
trilinear $\tilde{\tau}^*_R$--$\tilde{\tau}_L$--$H_1$ coupling
$A_\tau=|A_\tau|\, {\rm e}^{i\Phi_{A_\tau}}$ as well as the parameters
$\mu$ and $\tan\beta$. These seven real and positive parameters and
three {\em CP} violating phases completely determine the mass spectrum
and mixing in the stau and chargino/neutralino sectors:
\beq \label{eq:SUSY parameters}
\{|M_1|,\, M_2,\, |\mu|,\, \widetilde{m}_L,\, \widetilde{m}_R,\,
|A_\tau| ;\, \tan\beta\}\qquad {\rm and}\qquad \{\Phi_1,\, \Phi_\mu,\,
\Phi_{A_\tau}\}\,. 
\eeq

Through the analysis of the neutralino and chargino system, strategies
have been developed in great detail \cite{lcstudies, Choi:neutralino}
to determine the gaugino mass parameters $M_1$ and $M_2$ as well as
the higgsino parameter $\mu$. If $\tan\beta \gsim 10$ it is rather
difficult to accurately measure its value, since the neutralino and
chargino masses and mixing angles are not sensitive to $\tan\beta$ in
this case. On the other hand, the longitudinal tau polarization from
$\stau_1$ decay, which could be measured within about $5\%$ accuracy,
can be used to determine \cite{nojiri} high values of $\tan\beta$ with
an error of about $5\%$~\cite{Boos}. In Ref.~\cite{Boos}, it is also
shown that the measurement of both $\tilde{\tau}_i$ masses as well as
the $\tilde \tau_1^+ \tilde \tau^-_1$ production cross section in the
{\em CP} conserving case can determine the parameter $|A_{\tau}|$, if
$\mu$ is known from other measurements: Under favorable circumstances
an error of about $5\%$ in $\tan\beta$ and of about $5\%$ in
$m_{\tilde{\tau}_2}$ would result in the measurement of $|A_{\tau}|$
with an accuracy of about $8\%$. However, none of these observables is
sensitive to the {\em CP} violating phase $\phi_{\tilde \tau}$ in the
stau sector. On the other hand, this phase cannot be determined
independently of the other parameters in the $\tilde \tau$ mass
matrix.

We therefore re--examine the following observables in the stau and
neutralino sector:
\begin{enumerate}
\item The masses of the staus, $m_{\stau_{1,2}}$, and the masses of
the light neutralinos, $m_{\neu_{1,2}}$.

\item The cross sections of the neutralino pair production,
$\sigma(\tilde{\chi}^0_1 \tilde{\chi}^0_2)$, and of the stau
pair production, $\sigma(\tilde{\tau}_1 \tilde{\tau}_{1,2})$
with longitudinally polarized electron and positron beams.

\item The average polarizations of the second lightest neutralino and
tau leptons.

\item The spin/momentum correlations between the second lightest
neutralino and the tau lepton in the decay process $\tilde{\chi}^0_2
\rightarrow \tilde{\tau}_1 \tau$ and the spin/angular/momentum
correlations of the final two tau leptons.
\end{enumerate}
Only observables involving components of the $\tau$ spin orthogonal to
the $\tau$ momentum have a potential to probe $\phi_\stau$.

Obviously the detailed decay pattern depends on the stau and
neutralino mass spectrum.  Restricted to the case where the decay of
the $\neu_2$ into a stau is allowed, i.e. $m_{\neu_2} > m_{\stau_1}$,
the following two mass spectra are possible:
\begin{center}
\includegraphics{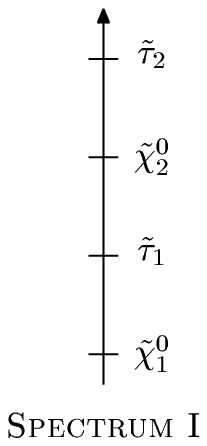}
\hspace{2cm}
\includegraphics{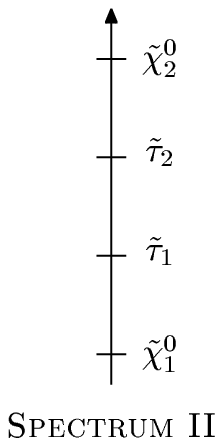}
\end{center}
These spectra result in different decay patterns:
\bea
\hbox{\sc Spectrum \phantom{I}I}: &&
\tilde{\tau}^\pm_2\rightarrow \tau^\pm \tilde{\chi}^0_{1,2},\quad
\tilde{\chi}^0_2\rightarrow \tau^\pm\tilde{\tau}^\mp_1, \quad
\tilde{\tau}^\pm_1\rightarrow \tau^\pm \tilde{\chi}^0_1\, ;
\label{eq:decay1} \\
\hbox{\sc Spectrum II}: &&
\tilde{\chi}^0_2\rightarrow \tau^\pm\tilde{\tau}^\mp_{1,2},\quad
\tilde{\tau}^\pm_2\rightarrow \tau^\pm \tilde{\chi}^0_1,\quad
\tilde{\tau}^\pm_1\rightarrow \tau^\pm \tilde{\chi}^0_1 \, .
\label{eq:decay2}
\eea
In the case of {\sc Spectrum I}, the production process $e^+e^-
\rightarrow \tilde{\chi}^0_2 \tilde{\chi}^0_2$ may also be possible if
$e^+e^- \rightarrow \tilde\tau_1^\pm \tilde\tau_2^\mp$ is accessible,
leading to events with four tau leptons and two lightest neutralinos
in the final state, from the sequential decay $\tilde{\chi}^0_2
\rightarrow \tau^+ \tau^- \tilde{\chi}^0_1$.

In the present work, we focus on \textsc{Spectrum I}\footnote{The
decay patterns of Eq.~(\ref{eq:decay2}) will lead to additional event
topologies from $\tilde{\chi}_1^0 \tilde{\chi}_2^0$ production,
requiring independent analyses.}. Considering the decays in
Eq.~(\ref{eq:decay1}), the production processes $e^+ e^- \rightarrow
\tilde{\chi}^0_1 \tilde{\chi}^0_2$ and $e^+ e^- \rightarrow
\tilde{\tau}^+_1 \tilde{\tau}^-_1$ can give rise to the same final
state with 2 $\tau$'s $+$ 2 LSP's, while the process $e^+ e^-
\rightarrow \tilde{\tau}^\pm_1 \tilde{\tau}^\mp_2$ could lead
eventually to (2 or 4) $\tau$'s $+$ 2 LSP's, or 2 $\tau$'s $+$ 2
$\nu_\tau$'s $+$ 2 LSP's if $\tilde{\tau}^\pm_2 \rightarrow
\tilde{\chi}^\pm_1 \nu_\tau$ and $\stau_1 \rightarrow \tilde
\chi_1^\pm$ are kinematically allowed. 

It is crucial to find some distinct features to disentangle these
reactions. First, we will assume that $\tilde \chi_1^0 \tilde
\chi_2^0$ production is studied at a beam energy where $\tilde
\tau_1^\pm \tilde \tau_2^\mp$ production is not accessible. This
leaves us with at most two competing processes; note that $\tilde
\chi_1^0 \tilde \chi_2^0$ production becomes possible at a lower
energy than $\tilde \tau_1$ pair production if $m_{\stau_1} > (
m_{\neu_1} + m_{\neu_2} ) /2$. If it is kinematically accessible,
$\tilde{\tau}^+_{1} \tilde{\tau}^-_1$ production tends to yield the
two $\tau$'s back to back, whereas $\tilde{\chi}^0_1 \tilde{\chi}^0_2$
production would have them more collinear, since they originate from
the same parent $\tilde{\chi}^0_2$. However, since above threshold
$\sigma(e^+ e^- \rightarrow \tilde \tau_1^+ \tilde \tau_1^-) \gg
\sigma(e^+ e^- \rightarrow \tilde \chi_1^0 \tilde \chi_2^0)$, angular
distributions will not be sufficient to suppress the $\tilde \tau_1$
pair background. In Secs.~5.3 and 6 we will therefore choose
parameters such that the $\tau$ energy distributions from these two
processes do not overlap.

\section{Stau and neutralino sector}
\label{sec:Mixing}

\subsection{Tau slepton mixing}

The mass-squared matrix of the stau is, in the basis $(\tilde \tau_L,
\, \tilde \tau_R)$ \cite{Nilles_Haber}:
\beq
\label{eq:stau mass matrix}
{\cal M}^{\,2}_{\tilde{\tau}}
=\left(\begin{array}{cc}
          \widetilde{m}_L^{\hspace{1mm}2}+m^2_\tau+D_L             &
          -m_\tau(A^*_\tau+\mu\tan\beta)\\[2mm]
          -m_\tau(A_\tau+\mu^*\tan\beta) &
          \widetilde{m}^{\hspace{1mm}2}_R+m^2_\tau+D_R
\end{array}\right)
=\left(\begin{array}{cc}
          m^2_{LL} &
          m^{2*}_{RL} \\[2mm]
          m^{2}_{RL} &
          m^2_{RR}
\end{array}\right)
\,,
\eeq
where $A_\tau$ is the complex trilinear coupling, $\mu$ is the complex
higgsino mass parameter, $\tan\beta$ is the ratio of the Higgs vacuum
expectation values, and $\widetilde{m}_L$ and $\widetilde{m}_R$ are
respectively the real $SU(2)$ doublet and singlet soft breaking mass
parameters. The $D$--terms are
\bea
D_L\!\!\!&=&\!\!\!\left(s^2_W-\frac{1}{2}\right)
\cos 2\beta\, m^2_Z \,,
\\ \nonumber
D_R\!\!\!&=&\!\!\!-s^2_W\cos 2\beta\, m^2_Z
\,,
\eea
where $s_W, c_W, t_W \equiv \sin\theta_W,\cos\theta_W,\tan\theta_W$.
Since $\cos2\beta = -(\tan^2\beta-1)/(\tan^2\beta+1)$, $D_L$ and $D_R$
become practically independent of $\tan\beta$ in the limit of large
$\tan\beta$. On the other hand, large $\tan\beta$ enhances the $\tau$
Yukawa coupling which in turn leads to substantial mixing between the
stau weak eigenstates. The off--diagonal entry $m^2_{RL}$ will then be
dominated by the contribution $\propto \mu^*$. It will therefore be
difficult to determine $A_\tau$ experimentally if $|A_\tau| \ll |\mu|
\tan\beta$.

The Hermitian matrix in Eq.~(\ref{eq:stau mass matrix}) is
diagonalized by a unitary matrix $U_\stau$, parameterized by a mixing
angle $\theta_{\tilde{\tau}}$ and a phase $\phi_{\tilde{\tau}}$:
\beq
\label{eq:Utau-para}
  U_\stau=
  \left(
  \begin{array}{ll}
    \cos\theta_\stau &
   -\sin\theta_\stau \, {\rm e}^{-i\phi_\stau} \\
    \sin\theta_\stau\, {\rm e}^{i\phi_\stau} &
    \phantom{-}\cos\theta_\stau  \\
  \end{array}
  \right),
\eeq
which leads to $U_\stau^\dagger {\cal M}^{\,2}_{\tilde{\tau}} U_\stau=
{\rm diag} (m_{\stau_1}^2,m_{\stau_2}^2)$ with the convention
$m_{\stau_2} \geq m_{\stau_1}$.

The physical masses are given by
\beq
m^2_{\tilde{\tau}_{1,2}}
   =\frac{1}{2}
   \left[ \thin m^2_{LL}+m^2_{RR}
     \mp\sqrt{(m^2_{LL}-m^2_{RR})^2+4|m^2_{RL}|^2} \thin
   \right]
   \,,
\eeq
and the stau mixing angle $\theta_\stau$ and the phase
$\phi_\stau$ are
\beq
\tan\theta_\stau = \frac{ m^2_{\tilde \tau_1} - m^2_{LL}
}{|m^2_{RL}|}, \qquad
\phi_\stau ={\rm arg}(m^2_{RL}) \,.
\eeq
Note that $-\pi/2 \leq \theta_{\tilde \tau} \leq 0$ in our convention,
whereas $\phi_{\tilde \tau}$ can take any value between $0$ and $2\pi$.

An important question is whether physical observables can
determine all the fundamental parameters in Eq.~(\ref{eq:stau mass matrix}).
Note that the mass parameters $\{m^2_{LL}$, $m^2_{RR}$,
$|m^2_{LR}|\}$ are conversely expressed by
\begin{eqnarray} \label{eq:stau mass parameters}
m^2_{LL}\!\!\!&=&\!\!\! \phantom{-}\frac{1}{2}\left[ \thin
m^2_{\stau_1} + m^2_{\stau_2} - (m^2_{\stau_2} - m^2_{\stau_1}) \cos
(2\thin\theta_\stau)  \right], \nonumber\\
m^2_{RR}\!\!\!&=&\!\!\! \phantom{-} \frac{1}{2}\left[ \thin
m^2_{\stau_1} + m^2_{\stau_2} + (m^2_{\stau_2} - m^2_{\stau_1}) \cos
(2\thin\theta_\stau) \right], \nonumber\\
|m^2_{RL}|\!\!\! &=& \!\!\! -\frac{1}{2} \thin( \thin m^2_{\stau_2} -
m^2_{\stau_1}) \sin(2\thin\theta_\stau) \,.
\end{eqnarray}
Since the cross sections for $\stau_i \thin \stau_j$ production depend
on the mixing angle $\theta_\stau$~\cite{nojiri,Bartl}, all the
quantities appearing on the r.h.s. of Eqs.~(\ref{eq:stau mass
parameters}) can be measured in $\stau$ pair production. In contrast,
the phase $\phi_\stau$ cannot be determined from {\em CP}--even
quantities such as the (polarized) production cross sections only. In
fact, it will affect $\stau$ production at $e^+e^-$ colliders and
their subsequent decay only in higher orders in perturbation theory,
unless $\stau_1$ and $\stau_2$ are so close in mass that $\stau_1 -
\stau_2$ oscillations become significant \cite{stauosc}.

\subsection{Neutralino mixing}

In the MSSM, the four neutralinos $\tilde{\chi}^0_i$ ($i=1,2,3,4$) are
mixtures of the neutral $U(1)$ and $SU(2)$ gauginos, $\widetilde{B}$
and $\widetilde{W}^3$, and the higgsinos, $\widetilde{H}^0_1$ and
$\widetilde{H}^0_2$~\cite{Nilles_Haber}. In the {\em CP}--violating
MSSM the neutralino mass matrix in the $(\widetilde{B},
\widetilde{W}^3, \widetilde{H}^0_1, \widetilde{H}^0_2)$ basis is
complex, given by
\begin{eqnarray}
{\cal M}_N=\left(\begin{array}{cccc}
   M_1  & 0   & -m_Z c_\beta s_W  &  m_Z s_\beta s_W \\
   0    & M_2 &  m_Z c_\beta c_W  & -m_Z s_\beta c_W \\
 -m_Z c_\beta s_W &  m_Z c_\beta c_W  & 0    & -\mu \\
  m_Z s_\beta s_W & -m_Z s_\beta c_W  & -\mu & 0
                 \end{array}\right)
                 \,.
\label{eq:neutralino mass matrix}
\end{eqnarray}
Since the matrix ${\cal M}_N$ is symmetric, a single unitary matrix
${N}$ is sufficient to relate the weak eigenstate basis with the mass
eigenstate basis of the Majorana fields $\tilde{\chi}^0_i$: $ {N}^*
{\cal M}_N {N}^\dagger = {\rm diag}(m_{\tilde{\chi}^0_1},
m_{\tilde{\chi}^0_2}, m_{\tilde{\chi}^0_3}, m_{\tilde{\chi}^0_4})$.

We note that in the limit of large $\tan\beta$, the gaugino--higgsino
mixing becomes almost independent of $\tan\beta$, unlike the stau
mixing. If $\tan\beta \gg 1$, measurements of neutralino masses and
mixings can therefore only yield a lower bound on this quantity.
Furthermore, the neutralino sector becomes independent of the phase
$\Phi_\mu$ in this limit.

\subsection{Interaction vertices}

In this section, we briefly summarize the interaction vertices
relevant for the production and decay of the staus and
neutralinos. The unitary matrices ${U}_\stau$ and ${N}$ determine,
respectively, the couplings of the stau mass eigenstates $\stau_{1,2}$
and the neutralinos $\neu_i$ to SM particles; both these matrices
appear in the $\stau$--$\neu$--$\tau$ couplings.

For the neutralino pair production with the selectron exchange
diagrams ignored, it is sufficient to consider the
neutralino--neutralino--$Z$ vertices:
\begin{eqnarray}
  \langle \tilde{\chi}^0_{iR}|Z|\tilde{\chi}^0_{jR}\rangle
 =-\langle \tilde{\chi}^0_{iL}|Z|\tilde{\chi}^0_{jL}\rangle^*
 = \frac{g}{2 \thin c_W}\left[N^*_{i3} N_{j3} - N^*_{i4} N_{j4}\right]
 \,.
\end{eqnarray}
The stau pair production processes $e^+ e^- \rightarrow \stau^\pm_i
\stau^\mp_j$ ($i,j=1,2$) proceed via $s-$channel $\gamma$ and $Z$
exchange. The relevant $\stau$--$\stau$--$\gamma$ and
$\stau$--$\stau$--$Z$ vertices are given by
\begin{eqnarray}
\langle \stau_i|\gamma|\stau_i\rangle = e,\quad \ \
\langle \stau_i|Z|\stau_j\rangle
       = -\frac{g}{c_W}\left[s^2_W \delta_{ij}
          - \frac{1}{2}(U_\stau)^*_{1i} (U_\stau)_{1j}\right]
          \,.
\end{eqnarray}

The decays of $\stau$ and $\tilde{\chi}^0_i$ involve both the stau and
neutralino mixing, described by the neutralino--stau--tau vertices
\begin{eqnarray}
&& \langle\neu_{i  }|\stau^-_a| \tau^- \rangle =
  -g \, \left[ \,(Q^{L}_{ i a})^*P_L + (Q^{R}_{ i a})^*P_R
  \right]\qquad (a=1,2\ \ {\rm and}\ \
  i=1,\cdots,4)
  \,,
\label{eq:nstautau}
\end{eqnarray}
where the left/right--handed couplings $Q^{L,R}$ are
\begin{eqnarray}
\label{eq:Qstau}
&& Q^L_{ i a} = \frac{N_{i 1}t_W}{\sqrt{2}}\,
     \left[\,- z_i (U_\stau)_{L a}
     + \sqrt{2} h_i (U_\stau)_{Ra}\right]\,,
   \nonumber\\
&& Q^R_{ i a} = \frac{N_{i1}^* t_W}{\sqrt{2}}\,
     \left[\, 2(U_\stau)_{Ra} + \sqrt{2} h^*_i (U_\stau)_{L a}\right]\,.
\end{eqnarray}
The coefficient $z_i$ ($h_i$) denotes the relative strength of the
left--handed gaugino (higgsino) contribution to the right--handed
gaugino contribution:
\begin{eqnarray}
z_i = \frac{N_{i1}t_W +N_{i2}}{N_{i1}t_W},
      \qquad
h_i = \frac{m_\tau}{\sqrt{2} \, m_W \cos\beta}\frac{N_{i3}}{
        N_{i1}t_W}
        \,.
\end{eqnarray}
Since in the large $\tan\beta$ limit the neutralino masses and mixing
are insensitive to $\tan\beta$, only the coefficient $h_i$ contains a
strong dependence on $\tan\beta$.

\section{Stau and neutralino pair production}
\label{sec:Production}

\subsection{Stau pair production}

The matrix element for $e^+ e^- \rightarrow \tilde{\tau}^-_i
\tilde{\tau}^+_j$ receives contributions from $\gamma$ and $Z$
exchange. Denoting by $\Theta_{\stau}$ the polar angle of
$\tilde{\tau}^-_i$ with respect to the $e^-$ beam direction, the
transition amplitudes for electron helicity $\sigma=\pm 1$ are
\begin{eqnarray}
{\cal T}
\left( e^+ e^- \rightarrow \tilde{\tau}^-_i\tilde{\tau}^+_j \right)
=- e^2 Z^\sigma_{ ij}\,\beta_{ij}\,\sin\Theta_{\stau} \,,
\end{eqnarray}
where $\beta_{ij} = \lambda^{1/2}(1, m^2_{\stau_i}/s,
m^2_{\stau_j}/s)$ with $\lambda(x, y, z) = x^2 + y^2 + z^2 -
2(xy+yz+zx)$. The vector chiral couplings are given by
\beq
\label{eq:Z-coupling}
 Z^\sigma_{ ij} = \delta_{ij}+ D_Z(s)
     \frac{s^2_W-(1-\sigma)/4}{c^2_W s^2_W}
     \left[\, s^2_W\, \delta_{ij}
           -\frac{1}{2} (U_\tau)^*_{1i} (U_\tau)_{1j}\right]
           \,,
\eeq
and the $Z-$boson propagator is $D_Z(s) = s / (s-m^2_Z+i
m_Z\Gamma_Z)$.  With the parameterization in Eq.~(\ref{eq:Utau-para}),
the cross section for $\stau_i \stau_i$ production depends on
$\cos^2\theta_\stau$. This is sufficient to determine $\theta_{\stau}$
uniquely, since it is constrained to lie between $-\pi/2$ and 0, as
remarked earlier.

\begin{figure}[htb]
\begin{center}
\includegraphics[scale=1.1]{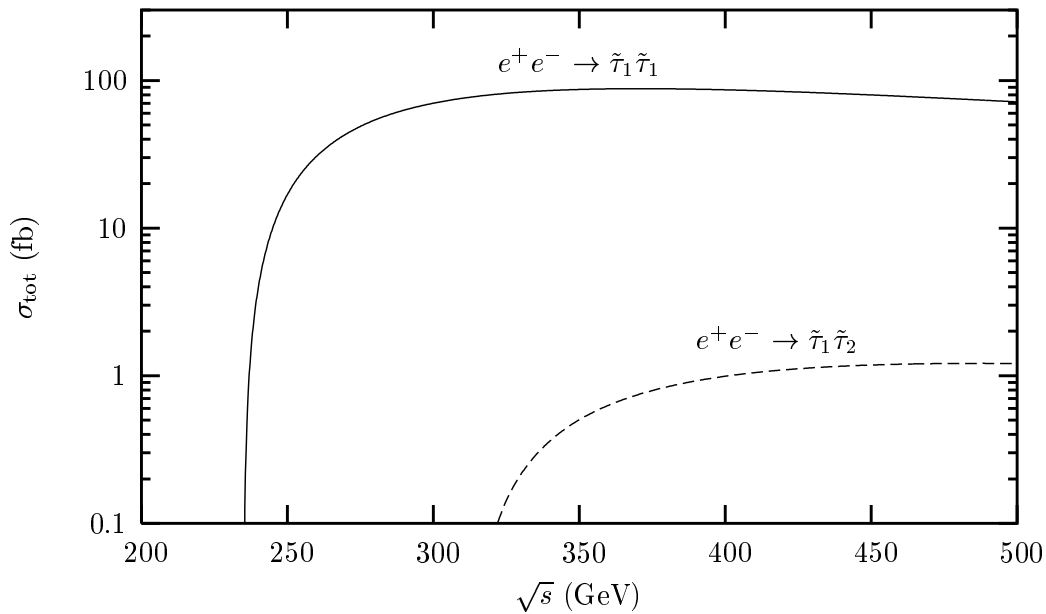}
\end{center}
\caption{\label{fig:stauPtot}
The total cross sections for $e^+ e^- \to \stau_1 \stau_1$ and
$e^+ e^- \to \stau_1 \stau_2 $ as a function of $\sqrt{s}$.
Parameters are set as explained in the text.}
\end{figure}

In Fig.~\ref{fig:stauPtot}, we present the total cross sections for
$e^+ e^- \to \stau_1 \stau_1$ and $e^+ e^- \to \stau_1 \stau_2$ with
unpolarized beams as a function of $\sqrt{s}$ for the following
parameters:
\bea
\label{eq:param-stau-P}
\widetilde{m}_{L} &=& 185~ \mathrm{GeV} , \quad
\widetilde{m}_{R} = 115~ \mathrm{GeV} , \quad
|A_\tau| = 1~ \mathrm{TeV},\quad
|\mu| = 200~ \mathrm{GeV},\\ \no
\Phi_\mu &=&0,\quad \Phi_{A} =0 \,.
\eea
The total cross section for $\stau_1\stau_1$ production is about 100
fb, large enough to probe the $\stau_1$ sector in detail. The
smallness of the off--diagonal couplings in Eq.~(\ref{eq:Z-coupling})
suppresses the production of $\stau_1^\pm\stau_2^\mp$ compared to that
of $\stau_1^\pm\stau_1^\mp$. As usual for P--wave processes, the cross
sections rise slowly near threshold, $\sigma_{\mathrm{thresh}} \propto
\beta^3$, rendering the $\stau$ mass determination through threshold
scans rather difficult
 
\subsection{Neutralino pair production}

As discussed earlier, we assume that first and second generation
sfermions are very heavy. The selectron exchange contributions to $e^+
e^- \rightarrow \tilde{\chi}^0_i \tilde{\chi}^0_j$ $(i,j=1\cdots 4)$
can therefore be ignored, leaving only the $s-$channel $Z$ exchange
diagram. The transition amplitudes can be expressed in terms of
generalized bilinear charges $\mathcal{Q}_{\alpha\beta}^{ij}$
\cite{Choi:neutralino}:
\begin{eqnarray} \label{eq:neutralino production amplitude}
\mathcal{T}\left( e^+ e^- \rightarrow \tilde{\chi}^0_i
\tilde{\chi}^0_j \right) = \frac{e^2}{s}\, \mathcal{Q}_{\alpha\beta}^{ij}
\left[\bar{v}(e^+)  \gamma_\mu P_\alpha  u(e^-)\right]
\left[\bar{u}(\tilde{\chi}^0_i) \gamma^\mu P_\beta
     v(\tilde{\chi}^0_j)\right]           \,.
\end{eqnarray}
They describe the neutralino production processes for polarized
electron and positron beams, neglecting the electron mass.  The first
lower index in $\mathcal{Q}_{\alpha\beta}^{ij}$ refers to the
chirality of the $e^\pm$ current, the second one to the chirality of
the neutralino current. These bilinear charges are
\begin{eqnarray} \label{eq:bilinear}
&& \mathcal{Q}_{LL}^{ij} = -\left( \mathcal{Q}_{RL}^{ij} \right)^* =
            \alpha_L D_Z {\cal Z}_{ij} \,,             \qquad
   \mathcal{Q}_{RL}^{ij} = -\left( \mathcal{Q}_{RR}^{ij} \right)^* =
            \alpha_R D_Z {\cal Z}_{ij}  \,,
\end{eqnarray}
where $\alpha_L = (s^2_W-1/2)/(s^2_W c^2_W)$ and $\alpha_R=1/c^2_W$.
The matrices ${\cal Z}_{ij}$ are defined as
\begin{eqnarray} \label{zdef}
{\cal Z}_{ij} = \frac{1}{2} \left( N_{i3} N^*_{j3} - N_{i4} N^*_{j4}
\right)  \,.
\end{eqnarray}
Eq.(\ref{zdef}) implies the {\em CP} relations ${\cal Z}_{ij} = {\cal
Z}^*_{ji}$, and hence $\mathcal{Q}_{\alpha\beta}^{ij} =
\left(\mathcal{Q}_{\alpha \beta}^{ji} \right)^*$ if the $Z$ width is
neglected.

It is known that polarized electron and positron beams are useful to
determine the wave--functions of the neutralinos
\cite{gudi,Choi:neutralino}. The electron and positron polarization
vectors are defined in the reference frame where the $z-$axis is in
the electron beam direction. We choose the electron and positron
polarization vectors as $P=(P_T,0,P_L)$ and $\bar{P}=(\bar{P}_T
\cos\eta,\bar{P}_T\sin\eta, -\bar{P}_L)$, respectively. The
differential cross section is then given by
\begin{eqnarray} \label{eq:diffx}
\frac{{\rm d}\sigma}{{\rm d}\Omega}\{ij\}
  =\frac{\alpha^2}{16\, s (1 + \delta_{ij})}\, \lambda^{1/2} \bigg[
     \left\{
     (1-P_L\bar{P}_L)+\xi\,(P_L-\bar{P}_L)\right\}\,\Sigma_U
  +P_T\bar{P}_T\cos\eta\,\Sigma_T\bigg]
  \,,
\end{eqnarray}
where $ \xi = (\alpha^2_R - \alpha^2_L) / (\alpha^2_R + \alpha^2_L) =
-0.147$ and the angular dependence of the coefficients $\Sigma_U$ and
$\Sigma_T$ is only from the polar angle $\Theta_{i}$ of the produced
neutralino $\neu_i$. Here $\lambda = [ 1 - (\mu_i+\mu_j)^2 ][ 1 -
(\mu_i-\mu_j)^2 ]$ with $\mu_i = m_{\neu_i}/\sqrt{s}$. Taking the
$Z$--boson propagator real by neglecting its width, the coefficients
read
\begin{eqnarray}  \label{eq:initial}
&& \Sigma_{U}=2 \thin D^2_Z (\alpha^2_R+\alpha^2_L)
           \left[\left\{1-(\mu^2_i - \mu^2_j)^2
                   +\lambda \thin \cos^2\!\Theta_i \right\}|Z_{ij}|^2
                   -4 \thin \mu_i\mu_j  \thin \real(Z^2_{ij}) \thin \right]
                   \,,
                  \nonumber\\
&& \Sigma_{T}=4\lambda\, D^2_Z  \thin \alpha_R \thin \alpha_L |Z_{ij}|^2
               \sin^2\!\Theta_i
               \,.
\end{eqnarray}
Note that the cross section is completely described by the two
quantities $|Z_{ij}|^2$ and $\real(Z^2_{ij})$ for each neutralino
pair production, if the polarization of the produced neutralinos is
ignored. Transversely polarized beams do not provide any independent
information on the neutralino mixing. In what follows, we therefore
assume only longitudinally polarized beams.

\begin{figure}
\begin{center}
\includegraphics[scale=1.2]{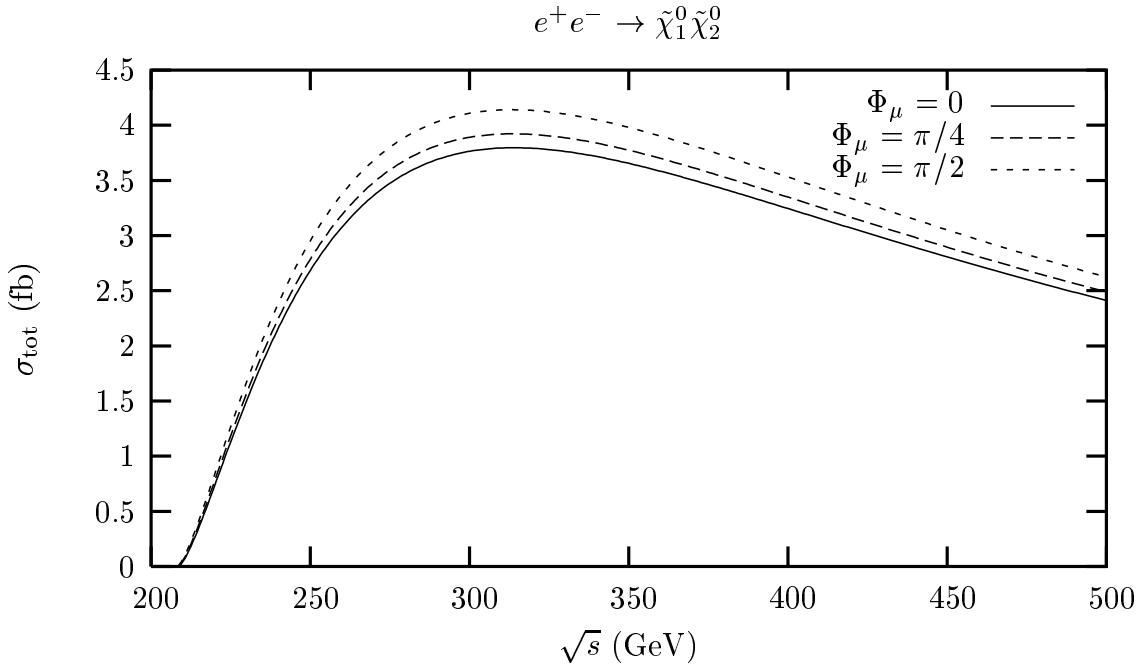}
\end{center}
\caption{\label{fig:ntotsig}
Total cross section for $e^+ e^- \to \neu_1\neu_2$ in fb.
We fix $\tan\beta=10$, $P_e = -0.8$, and  $\bar{P}_e =0.6$.}
\end{figure}

In order to probe the stau sector through the subsequent decay of
$\neu_2 \to \stau_1^\pm \tau^\mp$ following $e^+ e^- \to \neu_1
\neu_2$, the first question is whether the $\neu_1 \neu_2$ production
cross section is sufficiently large. In Fig.~\ref{fig:ntotsig}, we
show $\sigma_{tot}(e^+ e^- \to \neu_1 \neu_2)$ as a function of
$\sqrt{s}$ with $\tan\beta=10$ and $\Phi_1=0$. The beam polarizations
are set as $P_L = -0.8$ and $\bar{P}_L = 0.6$, which maximizes the
cross section if $|P_L| \leq 0.8$ and $|\bar P_L| \leq 0.6$. The same
choice of beam polarization minimizes the $\stau_1$ pair background,
if $\widetilde{m}_R < \widetilde{m}_L$ as expected in most SUSY
models. The gaugino mass unification relation $|M_1| = (5/3)t^2_W M_2$
with $\Phi_1 = 0$ is employed.  For $\Phi_\mu=0$, the parameters are
set as
\beq \label{eq:param-neuP}
M_1 = 85 ~\mathrm{GeV}, \quad |\mu|=200~\mathrm{GeV} \,.
\eeq
For $\Phi_\mu = \pi/4$ and $\pi/2$, parameters are chosen to yield the
same neutralino mass spectrum as the parameter set
Eq.~(\ref{eq:param-neuP}), i.e., 
\beq
m_{\neu_1} = 76 \pm 0.1 ~\mathrm{GeV}, \quad
m_{\neu_2} = 132 \pm 0.1 ~\mathrm{GeV}, \quad
m_{\neu_3} > 200~ \mathrm{GeV}\,.
\eeq
For given neutralino masses, a large phase $\Phi_\mu$ slightly
increases the total cross section.  Since the current expectation for
the annual luminosity of the future linear collider is about 1000
fb$^{-1}$, we have a few thousands events.

\subsection{Neutralino polarization vector
\label{sec:Neu-polarization}}

The neutralino polarization contains further information especially on
the chiral structure of the neutralinos. The polarization vector
$\vec{\cal P}^i=({\cal P}_T^i, {\cal P}_N^i,{\cal P}_L^i) $ of the
neutralino $\tilde{\chi}^0_i$ is defined in its rest frame. The
component ${\cal P}_L^i$ is parallel to the $\tilde{\chi}^0_i$ flight
direction in the $e^+ e^-$ c.m. frame, ${\cal P}_T^i$ is in the
production plane, which we take to be the $(x,z)$ plane, and ${\cal
P}_N^i$ is normal to the production plane, i.e. in $y$
direction. An explicit calculation gives
\begin{eqnarray} \label{eq:neutralino polarization}
{\cal P}^i_{L,T,N} = \frac{\xi\, (1-P_L \overline{P}_L)
                 + (P_L -\overline{P}_L)}{
                  (1-P_L \overline{P}_L)
                 + \xi\, (P_L -\overline{P}_L)}\cdot
                 \frac{\Delta_{L,T,N}}{\Sigma_{U}}          \,,
\end{eqnarray}
where $\xi$ and $\Sigma_U$ have been introduced in
Eq.(\ref{eq:diffx}). The coefficients $\Delta$ are
\begin{eqnarray} \label{Deltadef}
&& \Delta_L =\phantom{-} 4|D_Z|^2 (\alpha^2_R+\alpha^2_L)\cos\Theta_i
              \left[(1-\mu^2_i-\mu^2_j)\,|Z_{ij}|^2
               -2\mu_i\mu_j\,\real(Z^2_{ij})\right],
              \nonumber\\
&& \Delta_T=-4 |D_Z|^2 (\alpha^2_R+\alpha^2_L)\sin\Theta_i
              \left[(1-\mu^2_i+\mu^2_j)\mu_i\,|Z_{ij}|^2
               -(1+\mu^2_i-\mu^2_j)\mu_j\,\real(Z^2_{ij})\right],
              \nonumber\\
&& \Delta_N =-4 |D_Z|^2  (\alpha^2_R+\alpha^2_L)\sin\Theta_i
             \thin \lambda^{1/2}\mu_j \, \imag(Z^2_{ij})          \,.
\end{eqnarray}
Since $\xi = -0.147$ is small, a sizable neutralino polarization is
only possible in the presence of large beam polarization. The
measurement of the neutralino normal polarization is crucial to probe
$\imag (Z^2_{ij})$.

\begin{figure}[hbt]
\begin{center}
\includegraphics[scale=1.2]{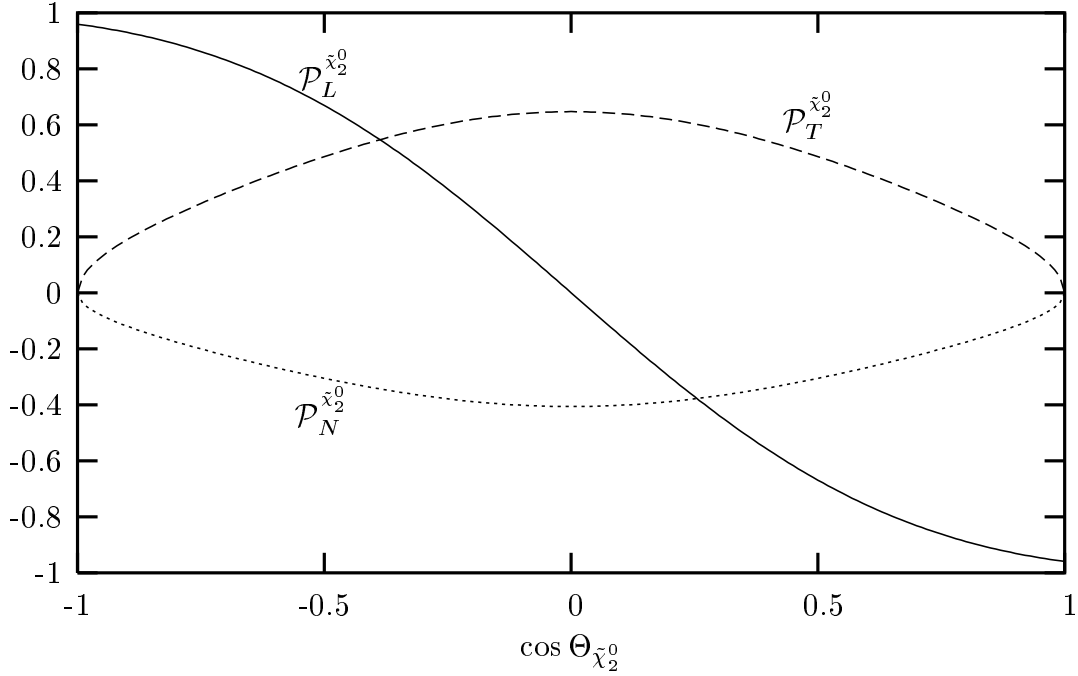}
\end{center}
\caption{\label{fig:Pi}
Polarization vector components of $\neu_2$ as a function of
$\cos\Theta_i$. We set $\sqrt{s} = 300$ GeV, $M_1=85$ GeV, $|\mu|=200$
GeV, $\tan\beta=10$, $\Phi_1 = \Phi_\mu = \pi/4$, $P_L = -0.8$ and
$\bar{P}_L =0.6$. }
\end{figure}

In Fig.~\ref{fig:Pi} we present the polarization vector of $\neu_2$ as
a function of $\cos\Theta_i$ with $\Phi_1=\Phi_\mu = \pi/4$.  The
sizable beam polarizations $P_L = -0.8$ and $\bar{P}_L = 0.6$ do
generate a substantial polarization of the neutralino. We will see in
the next Section that non--vanishing polarization of the neutralino
$\neu_2$ is essential for measuring the {\em CP}-violating phase
$\phi_\stau$ in the stau sector.

\section{Decay of the stau and neutralino}
\label{sec:Decay}

\subsection{Stau decays}

The decay distribution of the stau decay ${\stau}^\mp_a \rightarrow
\tau^\mp\tilde{\chi}^0_i$ and the polarization 4--vector of the final
tau lepton in the rest frame of the tau slepton are given by
\bea
\label{eq:stau-decay-amp}
&& \frac{d\Gamma_{\mp}}{d\Omega^*_1}
= \frac{g^2\, m_{\tilde{\tau}_a}\,\lambda^{1/2}} {64\, \pi^2}
\, (|Q^R_{ia}|^2 + |Q^L_{ia}|^2) \, \left[ 1 -
\frac{m^2_\tau}{m^2_{\tilde\tau_a} } - \frac {m^2_{\tilde\chi^0_i}}
{m^2_{\tilde\tau_a}} \, - 2 \frac{m_\tau m_{\tilde\chi^0_i} }
{m^2_{\tilde\tau_a} } \, {\cal A}^{ia}_T \right] \, , \\
&& {\cal P}^{\tau^\mp}_\mu = \mp \frac{ {\cal A}^{ia}_L }
{ 1 - 2m_\tau m_{\tilde{\chi}^0_i} / ( m^2_{\tilde\tau_a} - m^2_\tau
- m^2_{\tilde\chi^0_i} ) \, {\cal A}^{ia}_T } \, \left( \frac {
m_\tau\, q_{i\mu} } {k_1\cdot q_i} - \frac {k_{1\mu}} {m_\tau}
\right),
\label{taupol1}
\eea
where $d\Omega^*_1 = d\cos\theta^*_1 d\phi^*_1$ is the solid angle of
the tau lepton in the $\stau_a$ rest frame, $k_1$ and $q_i$ are the
4--momenta of the tau lepton $\tau^\mp$ and the neutralino
$\tilde{\chi}^0_i$, and the phase space factor $\lambda = \lambda(1,
m^2_\tau/m^2_{\stau_a}, m^2_{\tilde\chi^0_i} / m^2_{\stau_a} )$.
Expressions for the $Q^{L,R}_{ia}$ have been given in
Eq.~(\ref{eq:Qstau}). For the sake of convenience, we introduce three
tau polarization asymmetries:
\begin{eqnarray} \label{defA}
{\A}_L^{ia} = \frac {|Q^R_{ia}|^2-|Q^L_{ia}|^2}
 {|Q^R_{ia}|^2+|Q^L_{ia}|^2},  \quad
{\A}_T^{ia} = \frac{2 \thin \real(Q^R_{ia} Q^{L*}_{ia})}
 {|Q^R_{ia}|^2+|Q^L_{ia}|^2},  \quad
{\A}_N^{ia} = \frac{2 \thin \imag(Q^R_{ia} Q^{L*}_{ia})}
 {|Q^R_{ia}|^2+|Q^L_{ia}|^2} \,.
\end{eqnarray}

The average polarization of the $\tau$ lepton can be measured through
$\tau$ lepton decays within the detector
\cite{bullock,Davier}. Eq.(\ref{taupol1}) shows that it is purely
longitudinally polarized\footnote{The boost into the lab frame will in
general produce a small tranverse polarization; however, it is
suppressed by a factor $m_\tau/E_\tau$.}; its degree of polarization
is given by
\begin{eqnarray}
{\cal P}^{\tau^\mp}_L= \pm \frac{\lambda^{1/2}(m^2_{\tilde\tau_a}, m^2_\tau,
           m^2_{\tilde{\chi}^0_i})\, {\cal A}^{ia}_L}{
           m^2_{\tilde\tau_a}- m^2_\tau-m^2_{\tilde{\chi}^0_i}
                        - 2 m_\tau m_{\tilde{\chi}^0_i}\, {\cal A}^{ia}_T}
           \quad \rightarrow\quad \pm\, {\cal A}^{ia}_L\ \ {\rm for}\ \
           m_{\tilde{\tau}_a}\gg m_\tau \,.
\label{eq:stau-decay-pol}
\end{eqnarray}
In the rest frame of the stau lepton the decay distribution
(\ref{eq:stau-decay-amp}) is isotropic, as in all 2--body decays of
scalar particles. The degree of longitudinal polarization
(\ref{eq:stau-decay-pol}) is constant over phase space, depending only
on the couplings $Q^{L,R}_{ia}$. The magnitude of ${\cal
P}^{\tau^\pm}_L$ depends not only on the stau mixing but also on the
neutralino mixing as shown in Eq.~(\ref{eq:Qstau}). Since the gaugino
interaction with (s)fermions preserves chirality while the higgsino
interaction flips chirality, ${\cal P}^{\tau^\pm}_L$ is sensitive to
the $\tau$ Yukawa coupling~\cite{nojiri}. Note that $\tau$ mass
effects, which have been neglected in
refs.\cite{nojiri,Boos,bartltau}, introduce some dependence of ${\cal
P}_L^\tau$ on the phase $\phi_\stau$, via ${\cal A}_T^{ia}$. However,
this dependence is almost always too weak to allow a measurement of
this phase through $\stau_1 \rightarrow \neu_1 \tau$ decays. Not only
is $m_\tau m_{\neu_1} \ll m^2_{\stau_a} - m^2_{\neu_1} - m^2_\tau$
unless $m_{\neu_1}$ is very close to $m_{\stau_a}$, the coefficient
${\cal A}_T^{11}$ is usually also significantly smaller in magnitude
than 1.

Let us consider some limiting cases of the $\stau_1 \to\neu_1 \tau$
decay in the limit $m_{\stau_1} \gg m_\tau$, involving the couplings
$Q^{R,L}_{11}$. If the lightest neutralino is a pure Bino, the degree
of longitudinal polarization of the tau lepton becomes
\begin{eqnarray}
{\mathcal P}_\tau (\stau_1 \to \widetilde{B}\tau) =
\frac{4\sin^2\theta_\stau-\cos^2\theta_\stau}
{4\sin^2\theta_\stau+\cos^2\theta_\stau}    \,,
\end{eqnarray}
which has some dependence on the stau mixing angle. If the lighter
stau is right-handed, we have 
\beq 
{\mathcal P}_\tau (\stau_R \to \neu_1 \tau) = 
\frac{2 \thin t_W^2|N_{11}|^2 - Y_\tau^2 |N_{13}|^2} 
{2 \thin t_W^2|N_{11}|^2+Y_\tau^2|N_{13}|^2} \,,
\eeq 
where $Y_\tau = m_\tau / (\sqrt{2} m_W \cos\beta)$ is the $\tau$
Yukawa coupling divided by the $SU(2)$ gauge coupling. This will
deviate significantly from unity only if $\tan\beta$ is large, so that
$Y_\tau$ becomes comparable to the $U(1)_Y$ gauge coupling, {\em and}
the LSP has a significant higgsino component. Since in most models,
including the numerical examples we present below, the LSP is indeed
Bino--like and $\stau_1$ is dominantly $\stau_R$ (i.e. $\theta_\stau$
is near $-\pi/2$), the polarization of $\tau^-$ from $\stau_1^-$ decay
is usually quite close to $+1$, with little dependence on SUSY
parameters (within the ranges allowed by the model)
\cite{dproy}. Measuring this polarization can thus test this (large)
class of models, but is often not very useful for determining
parameters.

\subsection{Neutralino decays}

We are interested in the following neutralino production and decay
process:
\setlength{\unitlength}{1cm}
\begin{center}
\begin{picture}(7,2.2)(0,-2.2)
\put(.8,0)
 {\makebox(0,0)[t]{$e^+ + \,e^-\,\longrightarrow\,
                 \neu_j +\neu_i $}}
\put(3.6,-1.1)
 {\makebox(1,0)[l]{$\tau^\mp + \stau^\pm_a$}}
\put(6.6,-1.6)
 {\makebox(0,0)[t]{$\tau^\pm + \neu_1$}}

\put(2.4,-.6){\line(0,-1){.5}}
\put(2.393,-1.1){\vector(1,0){1}}

\put(4.7,-1.4){\line(0,-1){.5}}
\put(4.69,-1.9){\vector(1,0){1}}
\end{picture}
\end{center}
Since the spin-$1/2$ neutralino is polarized through its production as
described in Sec.~4.3, non--trivial spin correlations are generated
between the decaying neutralino and the tau lepton produced in the
first step of $\neu_i$ decay.  In order to describe the decay
distribution and tau polarization, we define a ``starred'' coordinate
system, where the $(x^*,z^*)$ plane is still the production plane of
the neutralino pair, but the neutralino $\neu_i$ momentum points along
the $z^*$ axis. The ``starred'' set of axes is thus related to the
coordinate system used in Sec.~4 through a rotation around the $y =
y^*$ axis by the production angle $\Theta_i$. In this coordinate
system, the polarization vector of the neutralino $\tilde{\chi}^0_i$
is $\vec{P}^i=({\cal P}^i_T, {\cal P}^i_N, {\cal P}^i_L)$ in the rest
frame of the neutralino. The expressions for the polarization
components are given in Eqs.~(\ref{eq:neutralino polarization}) and
(\ref{Deltadef}).

The angular distribution and the polarization vector of the $\tau$
lepton from $\neu_i$ decay is given in terms of the polar and
azimuthal angles $\theta^*_2$ and $\phi^*_2$ of the $\tau$ momentum
direction with respect to the neutralino momentum direction in the
rest frame of the neutralino by
\begin{eqnarray}
\label{chidec}
&& \frac{d\Gamma_{\mp}}{d\Omega^*_2}
     = \frac { g^2 \lambda^{1/2} \, E_\tau } {64\, \pi^2} \, \left(
     |Q^R_{ia}|^2 + |Q^L_{ia}|^2 \right) \,
      \left[1 + \mu_\tau {\A}^{ia}_T
      \pm \beta_\tau {\A}^{ia}_L \vec{\cal P}^i \cdot
     \hat{s}^*_3 \right] \,, \\
\label{eq:taupolA}
&& {\cal P}^{\tau^\mp}_L
= \frac{\pm \beta_\tau{\A}^{ia}_L
  + ( 1 + \mu_\tau {\A}^{ia}_T) (\vec{\mathcal{P}}^i \cdot
\hat{s}^*_3 ) } { 1 + \mu_\tau {\A}^{ia}_T \pm \beta_\tau
{\A}^{ia}_L (\vec{\cal P}^i \cdot \hat{s}^*_3) }  \,,
\nonumber\\
&& {\cal P}^{\tau^\mp}_T
=\frac{ ( \mu_\tau + {\A}^{ia}_T ) \, ( \vec{\cal P}^i \cdot
\hat{s}^*_1 ) - \beta_\tau {\A}^{ia}_N \, (\vec{\cal P}^i
\cdot \hat{s}^*_2) } { 1 + \mu_\tau{\A}^{ia}_T \pm \beta_\tau
{\A}^{ia}_L ( \vec{\cal P}^i \cdot \hat{s}^*_3 ) } \,,
\nonumber\\
&& {\cal P}^{\tau^\mp}_N
= \frac{ ( \mu_\tau + {\A}^{ia}_T) \, ( \vec{\cal P}^i \cdot
\hat{s}^*_2 ) + \beta_\tau {\A}^{ia}_N \, ( \vec{\cal P}^i
\cdot \hat{s}^*_1 ) } { 1 + \mu_\tau {\A}^{ia}_T \pm
\beta_\tau {\A}^{ia}_L ( \vec{\cal P}^i \cdot \hat{s}^*_3 ) } \,.
\end{eqnarray}
Here,
\beq \label{Etau}
E_\tau = \frac{m_{\neu_i}}{2} \left( 1 + \frac {m_\tau^2}
{m_{\neu_i}^2} - \frac {m_{\stau_a}^2} {m_{\neu_i}^2} \right)
\eeq
is the energy of the $\tau$ in the $\neu_i$ rest frame, $\mu_\tau =
m_\tau / E_\tau$, $\lambda^{1/2} \equiv \lambda^{1/2}(1, m^2_{\stau_a}
/ m^2_{\neu_i}, m^2_\tau / m^2_{\neu_i})$, and the tau lepton speed
$\beta_\tau$ is given by $\beta_\tau = \lambda^{1/2}/\left[1 -
(m^2_{\stau_a} - m^2_\tau) / m^2_{\neu_i} \right]$. Finally, the three
unit vectors $\hat{s}^*_{1,2,3}$ are defined by
\begin{eqnarray} \label{shatdef}
\hat{s}^*_1 &=& ( \cos\theta^*_2 \cos\phi^*_2\,, \cos\theta^*_2
                  \sin\phi^*_2\,, -\sin\theta^*_2 )\,,
                  \nonumber\\[1mm]
\hat{s}^*_2 &=& (-\sin\phi^*_2\,, \cos\phi^*_2\,, 0)\,,
                  \nonumber\\[1mm]
\hat{s}^*_3 &=& (\sin\theta^*_2 \cos\phi^*_2\,, \sin\theta^*_2
                  \sin\phi^*_2\,, \cos\theta^*_2)\,.
\end{eqnarray}
Note that ${\cal P}^{\tau^\mp}_T, {\cal P}^{\tau^\mp}_N, {\cal
P}^{\tau^\mp}_L$ are the polarization components of the $\tau$
polarization vector along the $\hat{s}^*_1, \hat{s}^*_2, \hat{s}^*_3$
directions, respectively. Combining the three polarization components
leads to the $\tau$ polarization 3--vector
\begin{eqnarray}
\vec{\cal P}^{\tau^\mp} =\frac{ (\mu_\tau+{\cal A}^{ia}_T)\, \vec{\cal P}^i
                          +\left[(1-\mu_\tau) (1-{\cal A}^{ia}_T)\,
                           (\vec{\cal P}^i\cdot \hat{s}^*_3)
                            \pm \beta_\tau {\cal A}^{ia}_L \right]
                           \hat{s}^*_3
                          - \beta_\tau {\cal A}^{ia}_N \,
                           (\vec{\cal P}^i\times \hat{s}^*_3)}{
  1+\mu_\tau{\A}^{ia}_T
 \pm\beta_\tau{\A}^{ia}_L (\vec{\cal P}^i\cdot \hat{s}^*_3)} \,.
\end{eqnarray}
The polarization 4--vector of the tau lepton in the neutralino rest
frame can be obtained by applying a Lorentz boost along the
$\hat{s}^*_3$ direction with the tau lepton speed $\beta_\tau$ to the
4--vector $(0,\vec{\cal P}^{\tau^\mp})$.

In the present work we will focus on the decay $\neu_2 \to \tau^\mp
\stau^\pm_1$. If the $\neu_2$ is unpolarized ($\vec{\cal P}^i=0$),
only the polarization asymmetry ${\A}_L^{21}$ defined in the first
Eq.(\ref{defA}) can be determined by measuring the longitudinal
polarization of the tau lepton. Some limiting cases are:
\begin{eqnarray}
\A_L^{21} (\neu_2 = \widetilde{W}^3) \!\!\!&=& \!\!\! -1 \,,\\ \nonumber
\A_L^{21} (\neu_2 = \widetilde{H}_1^0) \!\!\!&=&\!\!\! \cos
2\theta_\stau \,,\\ \nonumber
\A_L^{21} (\stau_1 = \stau_R)
\!\!\!&=&\!\!\! \frac{2 \thin |N_{21}|^2 \thin t_W^2
- Y_\tau^2 |N_{23}|^2}{2\thin |N_{21}|^2\thin t_W^2+Y_\tau^2 |N_{23}|^2}
\,.
\end{eqnarray}
The GUT relation $M_1 \simeq 0.5 M_2$ suppresses the value of
$|N_{21}|$ in most of the parameter space, but for $|\mu| > M_2$,
$|N_{23}|$ is also suppressed. Even a small $\stau_L$ component in
$\stau_1$ can therefore change $\A_L^{21}$ significantly, making it a
far more sensitive probe of SUSY parameters than $\A_L^{11}$.

If the polarization of the neutralino $\neu_2$ is sizable, which is
possible only with the longitudinal polarization of the $e^\pm$ beams,
$\A_T^{21}$ and $\A_N^{21}$ become measurable. The explicit expression of
the numerator of $\A_N^{21}$ is 
\bea
\label{eq:Im}
\Im m (Q^R_{21} Q^{L*}_{21})
\!\! &=&\!\! \phantom{-}
\sqrt{2}\, t_W Y_\tau \sin^2\theta_\stau
\Im m ( N^*_{21}N^*_{23} )
\\ \no
&&-({Y_\tau}/{\sqrt{2}}\,)\cos^2\theta_\stau
\Im m( N^*_{23} (N^*_{22}+N^*_{21}t_W ))
\\ \no &&
-t_W \sin\theta_\tau \cos\theta_\stau
\Im m(N^*_{21} (N^*_{22}+N^*_{21}t_W )e^{i\phi_\stau})
\\ \no &&
+Y_\tau^2 \sin\theta_\tau \cos\theta_\stau
\Im m( N^*_{23} N^*_{23} e^{-i\phi_\stau} )
\,.
\eea
$\Re e (Q^R_{21} Q^{L*}_{21})$ is the same with $\Im m$ replaced by
$\Re e$.  The stau {\em CP} violating phase $\phi_\stau$ is present in
the last two terms of Eq.~(\ref{eq:Im}). Note that these terms are
non--zero only in the presence of nontrivial $\stau_L - \stau_R$
mixing. This is not surprising, since $\phi_\stau$ is associated with
the off--diagonal elements of the stau mixing matrix
(\ref{eq:Utau-para}). These two terms thus increase with $\tan\beta$;
this increase is particularly rapid for the last term, due to the
factor $Y_\tau^2$, but this term is important only for $\tan\beta >
20$. On the other hand, if the $U(1)_Y$ gaugino mass is real, which is
true in our convention if gaugino mass unification also holds for
their phases, the first two terms in Eq.(\ref{eq:Im}) are proportional
to $\sin2\beta$, i.e. they become small as $\tan\beta$ becomes
large. Finally, recall that $\neu_{1,2}$ have to have significant
higgsino components in order to obtain a sizable $\neu_1 \neu_2$
production cross section, see Eq.(\ref{zdef}).\footnote{However, if
$\neu_1$ was higgsino--like, the $\neu_2 - \neu_1$ mass splitting
would be small, making the ordering $m_{\neu_1} < m_{\stau_1} <
m_{\neu_2}$ assumed in this analysis implausible.} Therefore the
necessary conditions for our process to be sensitive to the phase
$\phi_\stau$ are:
\begin{itemize}
\item sizable mixing $\theta_\stau$ in the stau sector, which is
helped by large $\tan\beta$;
\item sizable mixing between gauginos and higgsinos, which requires
$|\mu|$ to be not too much larger than $M_2$.
\end{itemize}

\subsection{Numerical results of tau polarization asymmetries}

Before presenting numerical results of the tau polarization
asymmetries for a sample parameter set, some discussions of
experimental issues are in order here. Since the final state consists
of two tau leptons with two LSP's, the first question is how to
determine which $\tau$ lepton comes from the primary $\neu_2$ decay.
For example, the negatively charged $\tau^-$ can be produced through
the following two decay channels:
\bea
\label{eq:Decay1}
\hbox{\textsc{Decay \phantom{I}I : }} &&
\neu_2\to \stau_1^+ \tau^- \quad \hbox{followed by}
\quad \stau_1^+ \to \neu_1 \tau^+,
\\ \label{eq:Decay2}
\hbox{\textsc{Decay II : }} &&
\neu_2\to \stau_1^- \tau^+ \quad \hbox{followed by}
\quad \stau_1^- \to \neu_1 \tau^- .
\eea
If these two processes are indistinguishable, a substantial
reduction of the efficiency is inevitable; recall that the (almost
purely longitudinal) polarization of the $\tau$ produced in $\stau$
decay depends only very weakly on $\phi_\stau$.

In the rest frame of the $\neu_2$, the $\tau^-$ energy for
\textsc{Decay I} is given by Eq.(\ref{Etau}) with $i=2,\, a=1$,
whereas for \textsc{Decay II} it is distributed over
\beq
\label{eq:EDecay}
\hbox{\textsc{Decay II : }}
E_{\tau^-} \in \left[ \gamma_{\stau_1} E^*_\tau - \beta_{\stau_1}
\gamma_{\stau_1} \left| \vec{p}^*_\tau \right|, \gamma_{\stau_1}
E^*_\tau + \beta_{\stau_1} \gamma_{\stau_1} \left| \vec{p}^*_\tau
\right| \right]\,.
\eeq
Here,
\beq \label{Estar}
E^*_\tau = \frac {m_{\stau_1}} {2} \left( 1 + \frac {m_\tau^2}
{m_{\stau_1}^2} - \frac{m_{\neu_1}^2} {m_{\stau_1}^2} \right)
\eeq
is the energy of the $\tau$ lepton from $\stau_1$ decay in the
$\stau_1$ rest frame, $\left| \vec{p}^*_\tau \right| = \sqrt{E^*_\tau
- m^2_\tau}$, $\gamma_{\stau_1} = E_{\stau_1}/m_{\stau_1}$,
$\beta_{\stau_1} \gamma_{\stau_1} = \sqrt{ E^2_{\stau_1} /
m^2_{\stau_1} - 1}$, with
\beq \label{Estau}
E_{\stau_1} = \frac {m_{\neu_2}} {2} \left( 1 - \frac {m_\tau^2}
{m_{\neu_2}^2} + \frac {m_{\stau_1}^2} {m_{\neu_2}^2} \right)
\eeq
being the energy of $\stau_1$ in the $\neu_2$ rest frame. The boost to
the lab frame will broaden the energy distributions from these two
decay chains. Nevertheless, for some choices of parameters these
ranges do not overlap. In such a situation the question which $\tau$
lepton originates from the $\neu_{2}$ decay can be answered using the
energies of two $\tau$ leptons. However, the neutrino(s) produced in
the decay of the $\tau$ limit(s) the measurement of the $\tau$ energy.
Especially in the decay modes $\tau \to \pi \nu$ and $\tau\to e \nu
\nu, \mu \nu\nu$, usually less than half of the $\tau$ energy is
visible. On the other hand, in the decays $\tau \to \rho \nu$ or $\tau
\to a_1 \nu$ the substantial mass of the $\rho$ or $a_1$ meson
enhances the visible energy of the $\tau$. Moreover, these decay modes
are known to be useful to measure the tau polarization
\cite{bullock,Davier}.

As remarked earlier, the situation is cleanest if the two $\tau$
energy distributions show little or no overlap even after the boost to
the lab frame. One safe case is when $m_{\stau_1}$ is close to either
$m_{\neu_1}$ or $m_{\neu_2}$. In this case the signal usually has one
rather hard and one rather soft $\tau$ so that the overlap of two
$\tau$ energy distributions is not serious. Moreover this signal is
easy to distinguish from the possible background process $e^+ e^- \to
\stau_1^\pm \stau_1^\mp$ followed by $\stau_1^\pm \to \tau^\pm \neu_1$
which tends to have either two soft tau's (if $m_\stau$ is close to
$m_{\neu_1}$), or two hard ones (if $m_\stau$ is close to
$m_{\neu_2}$).

The second issue is the measurement of $\tau$ polarization, which is
analyzed through its decay distributions with the decay modes $\tau
\to \pi\nu,~\rho\nu,~a_1\nu,~\mu\nu\bar{\nu},~e\nu\bar{\nu}$. The
$\tau \to \pi\nu$ decay mode is useful for determining the $\tau$
polarization only if the $\tau$ energy is known, which is the case in
$e^+e^- \to \tau^+ \tau^-$ production studied at LEP, but not in our
case. We consider only the $\rho\nu$ and $a_1\nu$ final states with
the combined branching ratio of about 34\%: The energy distribution of
$\rho$ or $a_1$ decay products can determine the $\rho$ or $a_1$
polarization which can specify, in turn, the $\tau$
polarization~\cite{bullock,Davier}. Unfortunately the efficiency of
the $\tau$ transverse polarization measurement is usually smaller than
that of the $\tau$ longitudinal polarization~\cite{Bartl:taupol}, and
is further reduced as the $\tau$ energy increases. Since the $\tau^-$
energy is approximately proportional to the mass difference between
$\neu_2$ and $\stau_1$, the following mass spectrum is best suited to
clearly probe $\A_{T,N}$:

\begin{center}
\includegraphics{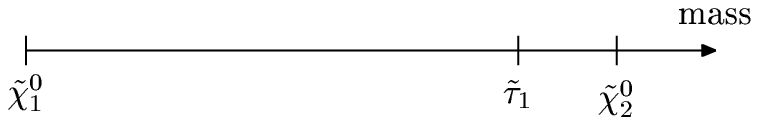}
\end{center}

The final issue is how to fix the parameters in the current situation
without a single signal of supersymmetric particles.  Eventually we
want to study the $\phi_\stau$-dependence of $\A^{21}_{L,T,N}$.
Varying the phase $\phi_\stau$ (through the phases $\Phi_\mu$ and/or
$\Phi_A$) while keeping all the other fundamental parameters the same
leads to different mass spectra in the neutralino and stau sectors.
This is not reasonable, since most likely these masses will be
measured earlier than the polarization observables we are considering.
The neutralino and stau sector is determined by the parameters of
Eq.(\ref{eq:SUSY parameters}), where the GUT relation $|M_1| = (5/3)
t_W^2 M_2$ is assumed.  As noted earlier, $\widetilde{m}_R$ is
expected to be smaller than $\widetilde{m}_L$ since $\stau_R$ has no
$SU(2)$ interactions. For definiteness, we fix
\beq \label{fixpar}
\tan\beta=10, \quad \Phi_1=0,\quad |A_\tau|=1~\mathrm{TeV}
\,.
\eeq
We vary all the other parameters. Considering the optimal scenario for
probing $\phi_\stau$ as discussed above, we fix the neutralino and stau
mass spectrum and stau mixing angle as follows:
\bea
\label{eq:mass-spectrum}
m_{\neu_1} &=& 80 \pm 0.5 ~\mathrm{GeV}, \quad 
m_{\neu_2} = 140 \pm 0.5 ~\mathrm{GeV}, \quad 
m_{\neu_3} = 225 \pm 5 ~\mathrm{GeV}, \quad \\ \no
m_{\stau_1} &=& 130 \pm 1 ~\mathrm{GeV}, \quad 
m_{\stau_2} = 210 \pm 1 ~\mathrm{GeV}, \quad 
\theta_\stau = -1.5 \pm 0.02 \,.  
\eea 
This constrains the mass parameters [in GeV]:
\beq
\label{eq:final-scenario}
M_1 \in [81.8,88.3], \quad
|\mu| \in [206,220], \quad
\widetilde{m}_R \in [122.5,128.5] , \quad
\widetilde{m}_L - \widetilde{m}_R \in [72,82.5] \,,
\eeq
while the {\em CP}--violating phase $\phi_\stau$ is completely
unconstrained.

\begin{figure}
\begin{center}
\includegraphics[scale=1.2]{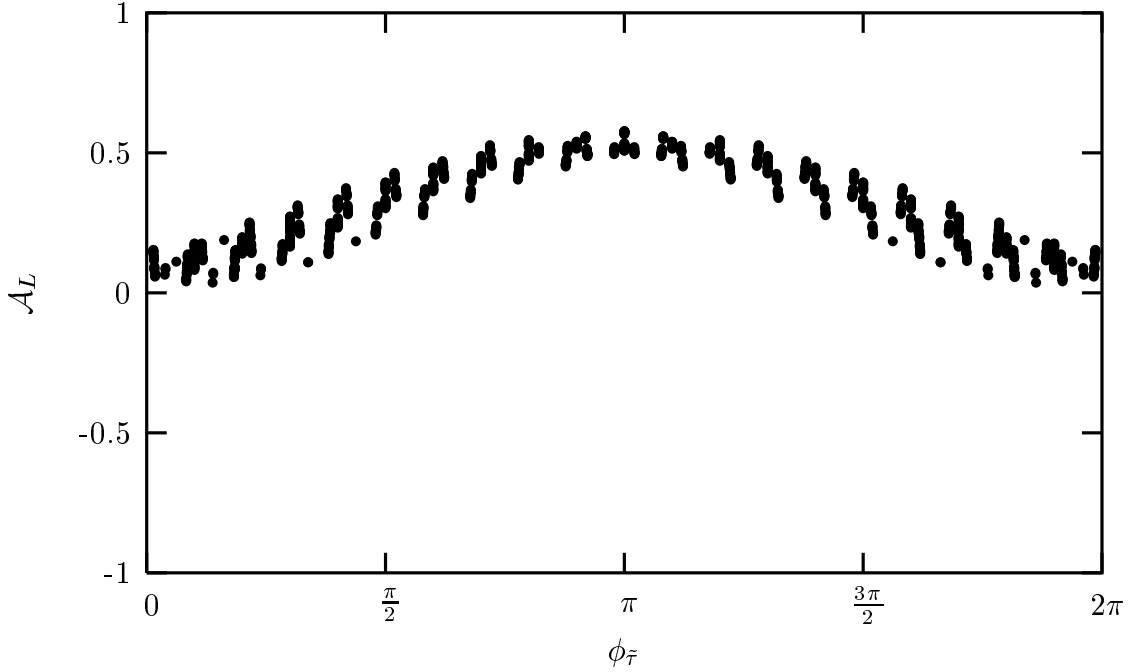}
\end{center}
\caption{\label{fig:AL}
The longitudinal polarization asymmetry of $\tau^-$ from
\textsc{Decay I}. Parameters are as in Eqs.(\ref{fixpar}) and
(\ref{eq:mass-spectrum}).}
\end{figure}

\begin{figure}
\begin{center}
\includegraphics[scale=1.03]{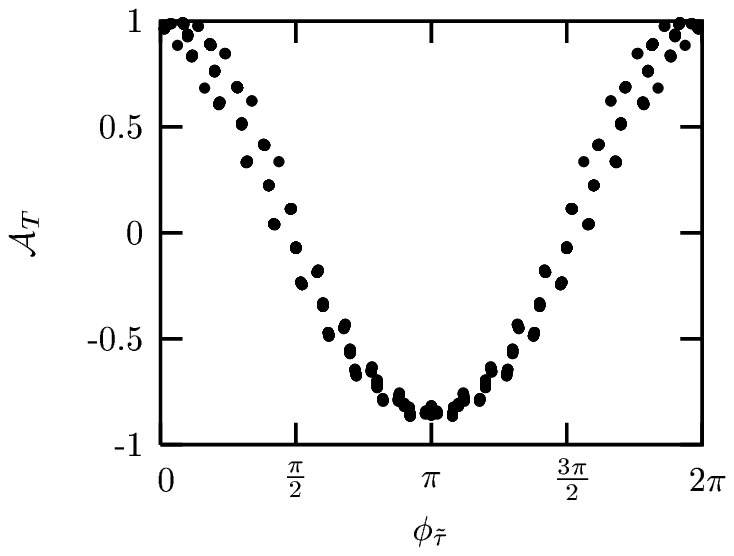}
\includegraphics[scale=1.03]{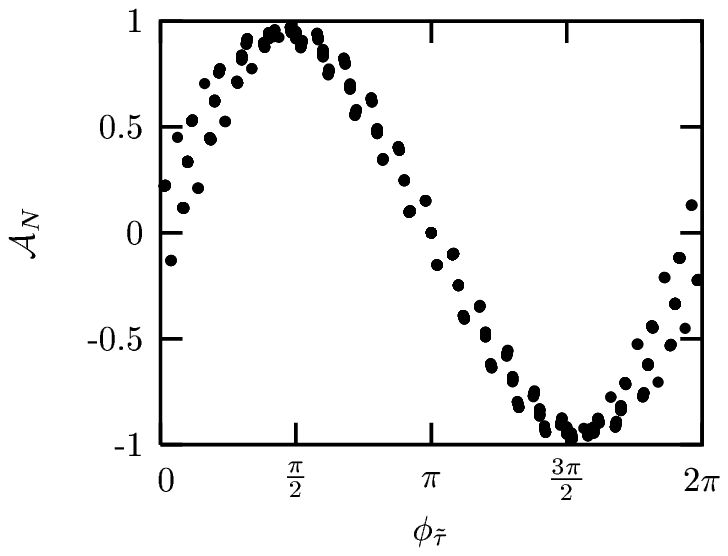}
\end{center}
\caption{\label{fig:ATN} 
The transverse and normal polarization asymmetries of $\tau^-$ from
\textsc{Decay I}. Parameters are as in Eqs.(\ref{fixpar}) and
(\ref{eq:mass-spectrum}).}
\end{figure}

Figure \ref{fig:AL} presents the longitudinal polarization asymmetry
of the $\tau^-$ lepton produced in \textsc{Decay I}, while
Fig.~\ref{fig:ATN} presents its transverse and normal polarization
asymmetries. Here ${\mathcal A}_{L,T,N} \equiv {\mathcal
A}_{L,T,N}^{21}$. The spread of the points for fixed $\phi_\stau$ is
attributed to our choice of SUSY scenario by fixing the neutralino and
stau mass spectra within finite error margins, rather than fixing the
fundamental SUSY parameters. As can be seen from Fig.~\ref{fig:AL},
this procedure introduces some dependence of $\A_L^{21}$ on
$\phi_\stau$, mostly through the change of the $\neu_2$ decomposition.
We had seen earlier that this quantity is very sensitive to various
SUSY parameters. However, the spread of the points is too large to
allow a good measurement of $\phi_\stau$. In contrast,
Fig.~\ref{fig:ATN} clearly shows that $\A_T^{21}$ and $\A_N^{21}$ are
quite sensitive to $\phi_\stau$. As expected from Eqs.(\ref{defA}),
they show complementary behavior: When $|{\mathcal A}_N| \simeq 1$
(particularly when $Q^R_{21}$ is almost real and $Q^L_{21}$ almost
imaginary or vice versa), ${\mathcal A}_T$ becomes minimized; when
$|{\mathcal A}_T| \simeq 1$ (particularly when both $Q^R_{21}$ and
$Q^L_{21}$ are almost either real or imaginary) ${\mathcal A}_N$
becomes minimized.

\section{Case studies}
\label{sec:Cases}

In the previous Section we saw that the polarization of the $\tau$
lepton produced in the primary $\neu_2$ decay $\neu_2 \to \stau^\pm
\tau^\mp$ depends sensitively on $\phi_\stau$ through the polarization
asymmetries $\A_{T,N}^{21}$. However, these quantities can be directly
extracted from the measurable $\tau$ polarization in the lab frame
only if the event can be reconstructed completely. In the case at hand
this would be true (up to possible discrete ambiguities) if the masses
of all participating superparticles were known, {\em and} if the
$\tau$ energies could be measured. Unfortunately, even in $\tau \to
\rho \nu, \, a_1 \nu$ decays a significant fraction of the $\tau$
energy will usually be carried away by the neutrino, making such an
approach impractical.

In this Section we therefore discuss the $\tau$ polarization in the
lab frame, as function of kinematical variables that are also defined
in the lab frame. To this end we have to boost the $\tau$ 4--momentum,
whose spatial component in the ``starred'' coordinate system points in
the direction of the unit vector $\hat{s}_3^*$ introduced in
Eq.(\ref{shatdef}), into the lab frame. In order to describe the
behavior of the $\tau$ lepton produced in the second step of the
$\neu_2$ cascade decay, we have to model $\stau_1$ decays in the
$\stau_1$ rest frame as described in Sec.~5.1, and again boost it into
the lab frame. Altogether we thus have to integrate over five angular
variables: the production angle $\Theta_2$ introduced in Sec.~4.2, and
the angles $\theta_1^*, \, \phi_1^*$ and $\theta_2^*, \, \phi_2^*$
describing $\stau_1 \to \tau \neu_1$ and $\neu_2 \to \stau_1 \tau$
decays, respectively. This is done using Monte Carlo methods.

We chose two points in the parameter space defined by
Eqs.(\ref{fixpar}) and (\ref{eq:mass-spectrum}). Both have
\beq \label{comsets}
\widetilde{m}_L = 205 \ {\rm GeV}\,, \quad
\widetilde{m}_R = 124 \ {\rm GeV}\,, \quad
|\mu| = 215 \ {\rm GeV}\,.
\eeq
\textsc{Set I} conserves {\em CP}:
\beq \label{set1}
\hbox{\textsc{Set \phantom{I}I : }} |M_1| = 87.5 \ {\rm GeV}\,, \quad
\Phi_\mu = 0\,, \quad \Phi_A = \pi \quad \quad \Rightarrow
\phi_\stau = \pi\,,
\eeq
while {\em CP} is violated for \textsc{Set II}:
\beq \label{set2}
\hbox{\textsc{Set II : }} |M_1| = 84.3 \ {\rm GeV}\,, \quad
\Phi_\mu = \frac{\pi}{2} = \Phi_A \quad \quad \Rightarrow \phi_\stau =
\frac{\pi}{2}\,. 
\eeq
Note that $|m_{RL}|$, and hence $\theta_\stau$, is the same in both
cases, whereas the phase $\phi_\stau$ changes.  We choose
center--of--mass energy $\sqrt{s} = 300$ GeV, where the signal cross
section is near its maximum, and take $P_L = -0.8, \, \overline{P}_L =
0.6$.

\begin{figure}[htb]
\begin{center}
\includegraphics[scale=.6]{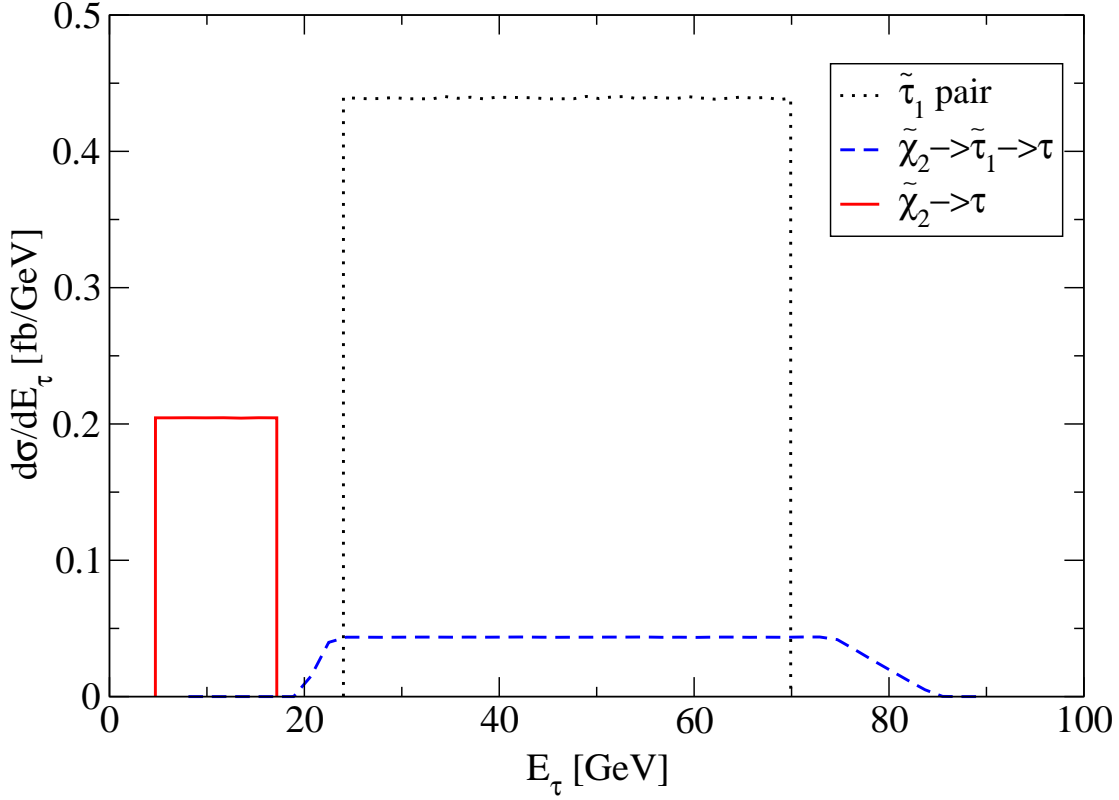}
\end{center}
\caption{\label{dsde}
Energy distribution of the ``soft'' (solid red) and ``hard'' (dashed
blue) $\tau$ lepton from $\neu_2$ decay, as well as for the $\tau$
lepton from $\stau_1$ pair production and decay (dotted
black). Parameters are as in Eqs.(\ref{fixpar}), (\ref{comsets}) and 
(\ref{set1}). }
\end{figure}

The resulting $\tau$ energy distributions are shown in
Fig.~\ref{dsde}. The (red) solid curve gives the energy distribution
for the ``soft'' $\tau$ from the primary $\neu_2$ decay, whereas the
dashed (blue) curve is for the ``hard'' $\tau$ lepton produced in the
second $\neu_2$ decay step. As advertised in Sec.~5.3, these two
distributions do not overlap.\footnote{At higher $\sqrt{s}$ some
overlap between these distributions does occur.} Moreover, the energy
distribution from $e^+ e^- \to \stau_1^+ \stau_1^- \to \tau^+ \tau^-
\neu_1 \neu_1$ (dotted black curve) is also well separated from that
of the ``soft'' $\tau$ lepton from $\neu_2$ decay. Note that this last
distribution is completely flat, even though Fig.~3, which uses very
similar parameters, shows that $\neu_2$ can be strongly polarized. The
energy of this $\tau$ lepton only depends on $\cos\theta_2^*$: it will
be maximal (minimal) if $\cos\theta_2^* = +1 \
(-1)$. Eq.(\ref{chidec}) shows that after integrating over $\phi_2^*$,
the $\neu_2$ decay distribution only depends on the longitudinal
$\neu_2$ polarization ${\cal P}_L^2$, which according to the first
Eq.(\ref{Deltadef}) is proportional to $\cos\Theta_2$. Integrating
over $\cos\Theta_2$ will therefore lead to a vanishing {\em average}
${\cal P}_L^2$, and hence to flat energy distributions of the primary
$\neu_2$ decay products. Since all masses are (essentially) the same
for our two sets, Fig.~\ref{dsde} is valid for both of them.

\begin{figure}[htb]
\begin{center}
\includegraphics[scale=.6]{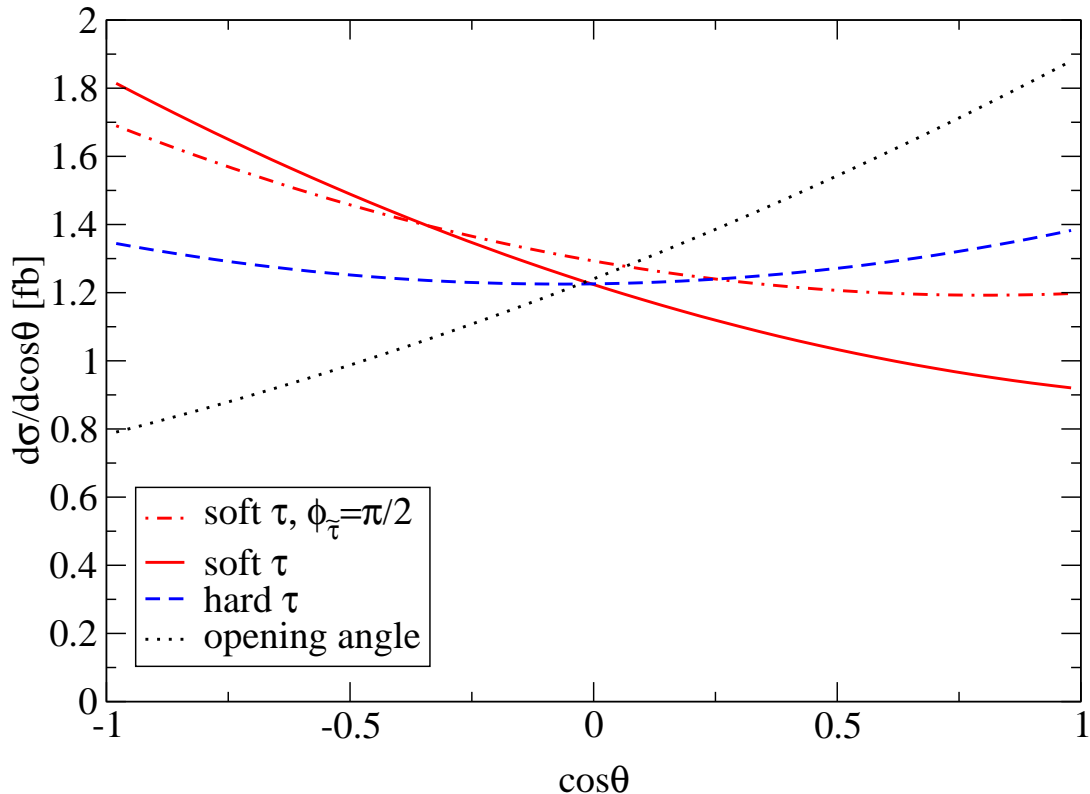}
\end{center}
\caption{\label{dsdcosth}
Angular distributions of the $\tau$ leptons from $\neu_2 \neu_1$
production and decay. Parameters are as described in the text.}
\end{figure}

Fig.~\ref{dsdcosth} shows angular distributions of the $\tau$ leptons
from \textsc{Decay I} of eq.(\ref{eq:EDecay}). The (red) solid and
(blue) dashed curves show the cross section as function of the cosine
of the angle between the $e^-$ beam direction and the direction of the
soft and hard $\tau$, respectively; the corresponding distributions
for \textsc{Decay II} can be obtained by sending $\cos\Theta
\rightarrow - \cos \Theta$. The ``soft'' $\tau$ shows quite a
pronounced forward--backward asymmetry. This can again be explained
from Eqs.(\ref{chidec}) and (\ref{Deltadef}). If $\neu_2$ goes in
forward direction, $\cos \Theta_2 > 0$, we have ${\cal P}_L^2 < 0$
(see Fig.~3). Since $\A_L^{21} > 0$, see Fig.~4, this means that the
soft $\tau$ will be emitted preferentially in the direction opposite
to that of $\neu_2$, i.e. in backward direction. On the other hand, if
$\cos \Theta_2 < 0$, we have ${\cal P}_L^2 > 0$, and the ``soft''
$\tau$ will be emitted preferentially collinear with $\neu_2$,
i.e. again in backward direction. The size of this effect depends on
$\A_L^{21}$: Fig.~4 shows a smaller $\A_L^{21}$ for $\phi_\stau =
\pi/2$ (\textsc{Set II}), leading to a less pronounced
forward--backward asymmetry, as illustrated by the (red) dot--dashed
curve. Finally, the (black) dotted curve shows the cross section as
function of the cosine of the opening angle between the two $\tau$
leptons. This distribution peaks at small angles, as expected from the
discussion at the end of Sec.~2. However, this peak is not very
pronounced, since the boost from the $\neu_2$ rest frame to the lab
frame is not very large. This distribution is the same for
\textsc{Decay I} and \textsc{Decay II}.

We know from the discussion of Sec.~5.3 that the main sensitivity to
$\phi_\stau$ comes from the components of the $\tau$ polarization that
are orthogonal to the $\tau$ 3--momentum. In principle, $\A_T^{21}$
and $\A_N^{21}$ are equally well suited to determine this
phase. However, observation of a {\em CP--} or {\em T--}odd quantity
would clearly be a more convincing proof of {\em CP} violation in the
stau and/or neutralino sector. Our choice of beam polarization implies
that the initial state is not {\em CP} self--conjugate. On the other
hand, since we are working in Born approximation and are neglecting
finite particle width effects, we can replace the {\em T}
transformation by the so--called naive $\widetilde T$ transformation,
which reverses the directions of all 3--momenta and spins, but does
{\em not} exchange initial and final state. Recall, however, that we
define the $e^-$ beam to go in $+z$ direction, whereas the transverse
component of the $\neu_2$ 3--momentum defines the $+x$ direction. In
our coordinate system a $\widetilde T$ transformation therefore
amounts to only changing the $y$ components of all 3--momenta and
spins; recall that the $y$ axis is the same in the original coordinate
system of Sec.~4 and the ``starred'' system introduced in
Sec.~5.2. Practically speaking, the $\widetilde T$ conjugate of some
kinematic configuration can therefore be obtained by simply sending
$\phi_1^* \rightarrow - \phi_1^*$ and $\phi_2^* \rightarrow -
\phi_2^*$. The existence of $\widetilde T$, and hence {\em CP},
violation is established if some observable takes {\em different}
values for some configuration and the $\widetilde T$ conjugate
one. Note that these two configurations result in the same $\tau$
energies; the angular variables whose distributions are shown in
Fig.~\ref{dsdcosth} also remain unchanged.

The simplest such observables involve the triple product of three
momentum or spin vectors. The triple product of the momenta of the
final--state leptons with the incoming $e^-$ momentum has been studied
in refs.\cite{oldcp}. This observable is sensitive to {\em CP}
violation in the neutralino sector, but is {\em not} sensitive to
$\phi_\stau$: We saw above that the $\tau$ energy distribution only
depends on $\A_L^{21}$, not on $\A_{T,N}^{21}$. Here we therefore
study the component of the ``soft'' $\tau$ polarization that is normal
either to the ``event'' plane defined by the $e^-$ beam and the
3--momentum of this $\tau$, or normal to the ``$\tau\tau$'' plane
spanned by the 3--momenta of the two $\tau$'s in the final state. In
both cases we also analyze the ``transverse'' component of the $\tau$
polarization of the ``soft'' $\tau$ leptons that lies in this
plane.

\begin{figure}[htb]
\begin{center}
\includegraphics[scale=.6]{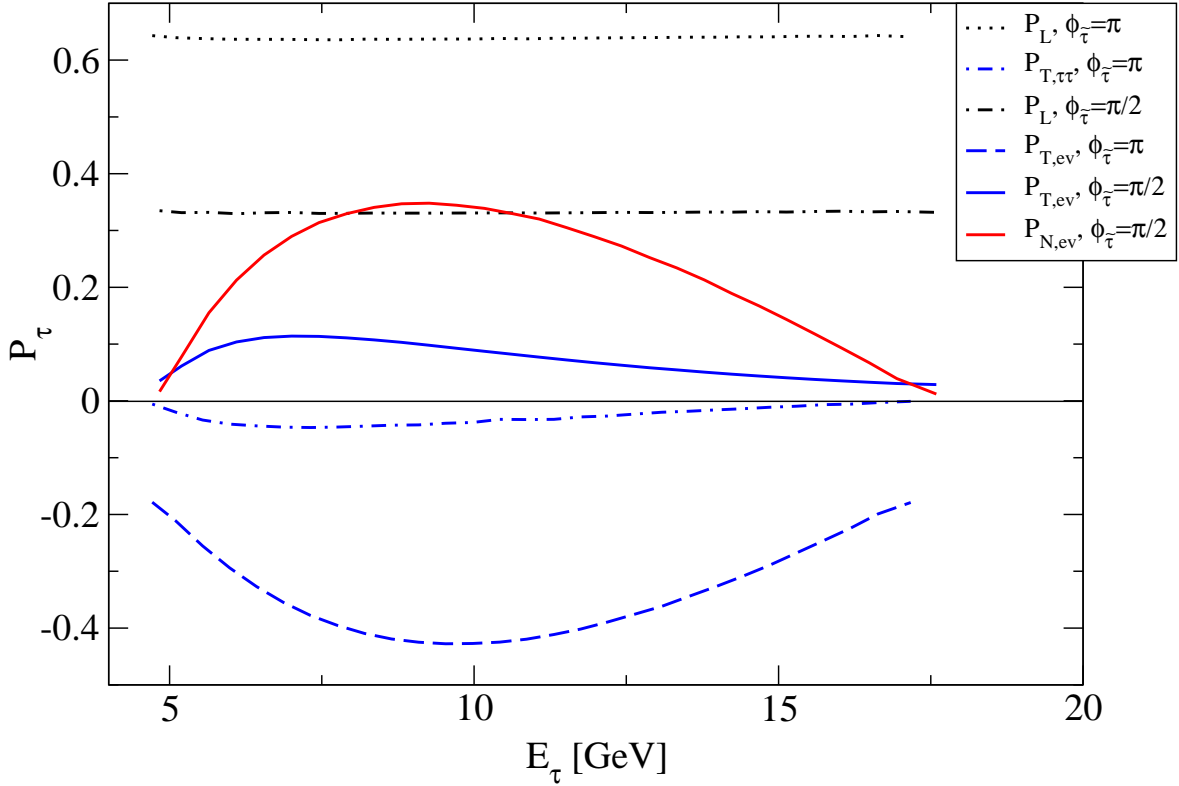}
\end{center}
\caption{\label{pole}
Energy dependence of the components of the polarization vector of the
$\tau^-$ produced in \textsc{Decay I}. The transverse $T$ and normal
$N$ components are defined either with respect to the plane spanned by
the $e^-$ and $\tau$ 3--momenta (subscript ``ev''), or with respect to
the plane spanned by the 3--momenta of the two $\tau$ leptons in the
final state (subscript ``$\tau \tau$''). Phase $\phi_\stau = \pi \
(\pi/2)$ refers to parameter \textsc{Set I} (\textsc{Set II}).}
\end{figure}

The dependence of the polarization of the soft $\tau$ lepton on its
energy is shown in Fig.~\ref{pole} for \textsc{Decay I}. The (black)
dotted and dash--doubledotted curves show the longitudinal
polarizations for \textsc{Set I} and \textsc{Set II}, respectively. We
see that it is essentially independent of $E_\tau$ after integrating
over the production angle $\Theta_2$. This component is
boost--invariant, up to terms of order $m_\tau^2$. Its average value
can thus be computed from Eqs.(\ref{chidec}) and (\ref{eq:taupolA}):
\beq \label{polav}
\langle {\cal P}_L^{\tau^\mp} \rangle \simeq \pm \frac {\beta_\tau \A_L^{21} }
{1 + \mu_\tau \A_T^{21} }\,,
\eeq
which well describes the numerical results of Fig.~\ref{pole}. Note in
particular that this component of the polarization has opposite sign
for \textsc{Decay II}, where the ``soft'' $\tau$ is positively
charged.

The (blue) dashed and lower solid curves show the transverse
polarization in the ``event'' plane for \textsc{Set I} and \textsc{Set
II}, respectively. It clearly mirrors the behavior of $\A_T^{21}$
shown in Fig.~5, being sizable and negative for $\phi_\stau = \pi$,
but small for $\phi_\stau = \pi/2$. It has some dependence on the
$\tau$ energy, reaching its maximal absolute value for some
intermediate $E_\tau$. This corresponds to small values of
$\cos\theta_2^*$, i.e. $|\sin\theta_2^*| \simeq 1$, which in turn
maximizes the product $\vec{\cal P}^2 \cdot \hat{s}_1^*$ that
multiplies $\A_T^{21}$ in the second eq.(\ref{eq:taupolA}); recall
from Fig.~3 that ${\cal P}_L^2$ is the potentially biggest component
of $\vec{\cal P}^2$. In contrast, the (blue) dot--dashed curve shows
that the transverse polarization in the ``$\tau \tau$'' plane is small
even for $\phi_\stau = \pi$. We suspect that such a small
polarization, of less than 5\%, will be very difficult to measure.

The upper (red) solid curve in Fig.~\ref{pole} shows the $\tau$
polarization normal to the ``event'' plane for \textsc{Set II}; since
this is a {\em T}--odd quantity, it vanishes identically for
\textsc{Set I}.  It again tracks the behavior shown in Fig.~5, being
large and positive for $\phi_\stau = \pi/2$. Also like ${\cal
P}_{T,{\rm ev}}^\tau$, it shows a pronounced extremum at intermediate
values of $E_\tau$. Note that the product $\vec{\cal P}^2 \cdot
\hat{s}_1^*$, which is maximized in this region of phase space, now
multiplies the {\em CP}--odd quantity $\A_N^{21}$ in the last
Eq.(\ref{eq:taupolA}). The $\tau$ polarization normal to the
``$\tau\tau$'' plane (not shown) is always much smaller than that
normal to the ``event'' plane.

\begin{figure}[htb]
\begin{center}
\includegraphics[scale=.6]{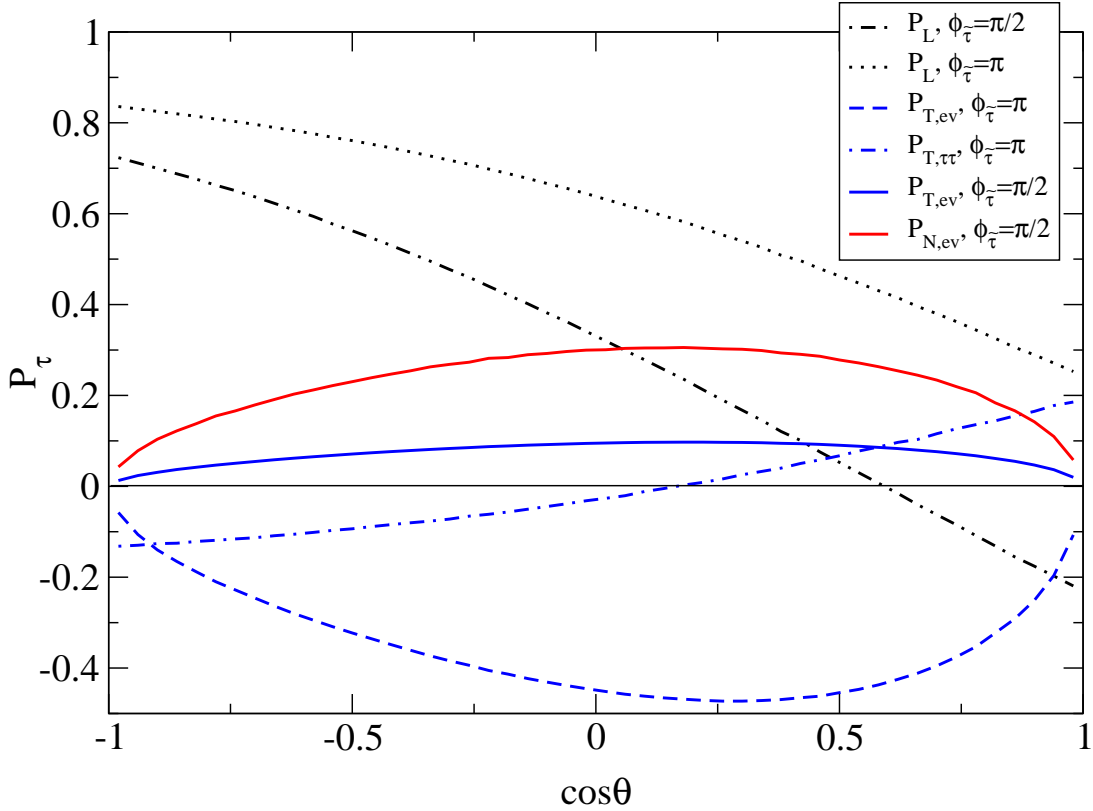}
\end{center}
\caption{\label{polth}
Dependence of the components of the polarization vector of the
$\tau^-$ produced in \textsc{Decay I} on the cosine of the angle
between this $\tau$ and the $e^-$ beam. Parameters and notation are as
in Fig.\ref{pole}.}
\end{figure}

Fig.~\ref{polth} shows the same $\tau$ polarization components as in
Fig.~\ref{pole}, as function of the cosine of the angle between the
incoming $e^-$ and outgoing $\tau$ 3--momenta. We see that the
longitudinal $\tau$ polarization depends quite strongly on this
angle. A value of $\cos \Theta_\tau$ near $-1$ is easiest to achieve
if $\cos \Theta_2 \simeq -1$ and $\cos \theta_2^* \simeq +1$, which
results in $\vec{\cal P}^2 \cdot \hat{s}_3^* > 0$ after integrating
over $\phi_2^*$, so that both terms in the numerator of the first
eq.(\ref{eq:taupolA}) are positive. In contrast, $\cos \Theta_\tau
\simeq 1$ is most easily achievable if $\cos \Theta_2 \simeq \cos
\theta_2^* \simeq +1$, which gives a negative product $\vec{\cal P}^2
\cdot \hat{s}_3^*$. For parameter \textsc{Set II}, where $\A_L^{21}$
is smaller, this even leads to negative ${\cal P}^\tau_L$ in the
forward direction. As before, ${\cal P}_L^\tau$ has to be reversed for
\textsc{Decay II}.

The overall behavior of the transverse and normal components defined
w.r.t. the ``event'' plane again follows the behavior of $\A_T^{21}$
and $\A_N^{21}$, respectively, as displayed in Fig.~5. We saw above
that these components reach their maximal values if $|\sin\theta_2^*|
\simeq 1$, which implies small $\cos\theta_2^*$ and also  the lab
system variable $|\cos \Theta_\tau|$ well away from unity. On the other
hand, we now see that the tranverse $\tau$ polarization defined
w.r.t. the ``$\tau \tau$'' plane can reach up to 20\% for parameter
\textsc{Set I}, which however is still well below the maximal value of
$|{\cal P}_{T,{\rm ev}}^\tau|$.

Finally, we also investigated $\vec{\cal P}^\tau$ as function of the
opening angle between the two $\tau$ leptons in the final state. In
all cases we find a very weak dependence on this angle. Note that this
angle is independent of the production angle $\Theta_2$, and only
weakly dependent on the $\neu_2$ decay angle $\theta_2^*$. We saw above
that these two variables largely determine the $\tau$
polarization. Integrating over them thus essentially reduces these
polarizations to their average values.

\section{Summary and Conclusions}
\label{sec:Conclusions}

In this paper we have investigated associated $\neu_1 \neu_2$
production followed by the two--step decay $\neu_2 \to \stau^\pm
\tau^\mp \to \neu_1 \tau^+ \tau^-$. We have seen that the components
of the $\tau$ lepton produced in the first step of this decay that are
orthogonal to the $\tau$ momentum are very sensitive to the {\em
CP}--odd phase $\phi_\stau$ in the stau sector. Much of this
sensitivity survives after boosting into the lab frame; the most
sensitive region of phase space involves intermediate values of both
the $\tau$ energy and its angle with respect to the beam direction. In
particular, we found a {\em CP}--violating normal component in excess
of 30\% in certain regions of phase space, if $\phi_\stau =
\pi/2$. Strong (longitudinal) beam polarization is crucial to find
such large {\em CP}--violating effects.

Of course, this result depends on the assumptions we made. To begin
with, we assumed an ``inverted hierarchy'' where the first and second
generation sfermions are very heavy. This is a conservative assumption
in the sense that it reduces our signal cross section by about two
orders of magnitude. On the other hand, it removes the otherwise very
stringent constraints on {\em CP} violation in the neutralino sector,
in particular on the phase $\Phi_\mu$ of $\mu$. It also implies that
the decay mode we are investigating has branching ratio near 100\% if
$\stau_1$ lies in between the two neutralinos. Here we analyzed a
situation with relatively small neutralino masses, where competing
2--body decay processes $\neu_2 \to \neu_1 Z$ or $\neu_2 \to \neu_1 h$
are not open; however, for gaugino--like neutralinos (which generally
have sizable mass splittings) these competing decays have rather small
branching ratios even if they are allowed.

A second important assumption is that $\stau_L - \stau_R$ mixing
should not be too small. This should be clear, since for vanishing
mixing the phase $\phi_\stau$ looses its physical meaning. We found
that the rather moderate choice $\tan\beta = 10$ is quite sufficient
for this purpose. This value of $\tan\beta$ is already so large that
the neutralino masses and couplings show relatively little sensitivity
to the phase $\Phi_\mu$. Our CP--odd observable is therefore indeed
mostly attributable to the stau sector.

We also assumed that $\stau_1$ is quite close in mass to
$\neu_2$. This makes it easy to decide on an event--by--event basis
whether $\neu_2$ decays into a positive or negative $\stau_1$, or
equivalently, which of the two $\tau$ leptons in the final state is
produced in the first step of $\neu_2$ decay; note that only this
lepton can have sizable transverse and normal polarization
components. However, even if no distinction between the two $\neu_2$
decay chains was possible, the necessary averaging would only reduce
the transverse and normal polarization asymmetries by a factor of 2.

Within the framework of inverted hierarchy models, our main assumption
is thus that $\neu_2 \rightarrow \stau_1 \tau$ 2--body decays are open
but decays into $\stau_2$ are not. If this second decay mode was also
open, more decay chains would need to be investigated; if they cannot
be distinguished experimentally, one may have to average over them,
which could lead to further degradation of the $\tau$
polarization. However, in most SUSY models the region of parameter
space where $m_{\stau_2} < m_{\neu_2}$ is rather small. On the other
hand, $m_{\stau_1} > m_{\neu_2}$ seems quite feasible. If the
competing $\neu_2$ 2--body decays are also closed, $\neu_2 \rightarrow
\tau^+ \tau^- \neu_1$ would still have a large, often dominant,
branching ratio; in many cases virtual $\stau$ exchange would give
significant contributions \cite{neutdec}. We therefore believe that in
this case sizable polarizations that are sensitive to $\phi_\stau$ can
again be found. If the competing $\neu_2$ 2--body decays are open but
the decay into $\stau_1$ is not, $\neu_2$ decays would not be a good
probe of $\phi_\stau$, since the final state of interest would then
only receive a very small contribution from virtual $\stau$
exchange. However, in this case one may still be able to extract this
information from analyses of the decays of heavier neutralinos, which
of course would require a somewhat higher beam energy.

We therefore conclude that neutralino decays into $\tau$ pairs offer a
good, indeed probably the best, possibility to probe {\em CP}
violation in the stau sector at $e^+e^-$ colliders.

\subsection*{Acknowledgements}
This work was partially supported by the KOSEF--DFG Joint Research
Project No. 20015-111-02-2. In addition, the work of SYC was supported
in part by the Korea Research Foundation through the grant
KRF-2002-070-C00022 and in part by KOSEF through CHEP at Kyungpook
National University. The work of JS is supported by the faculty
research fund of Konkuk University in 2003, and partly by Grant
No. R02-2003-000-10050-0 from BR of the KOSEF. The work of MD and BG
was partially supported by the Deutsche Forschungsgemeinschaft, grant
Dr 263.


\end{document}